\documentclass[aps,prb,twocolumn,superscriptaddress]{revtex4-1}

\usepackage{hyperref} 
\usepackage{graphicx}
\usepackage{amsmath}
\usepackage{amssymb}

\usepackage{standalone}
\usepackage{subfiles}

\usepackage{pdfpages}

\makeatletter
\AtBeginDocument{\let\LS@rot\@undefined}
\makeatother

\usepackage{amsmath}
\usepackage{pifont}
\usepackage{amssymb}
\usepackage{latexsym}
\usepackage{dsfont }
\usepackage{graphicx}
\usepackage{float}
\usepackage{color}
\usepackage{multirow}
\usepackage{verbatim} 

\newcommand{\SRO}{Sr$_2$RuO$_4$}

\def\bk{{\bold{k}}}

\def\m1{{^{-1}}}

\begin{document}


\title{Superconducting order parameter of Sr$_2$RuO$_4$: a microscopic perspective}

\author{Aline Ramires}
\affiliation{Max Planck Institute for the Physics of Complex Systems, Dresden, 01187, Germany}
\affiliation{ICTP-SAIFR, International Centre for Theoretical Physics - South American Institute for Fundamental Research, S\~{a}o Paulo, SP, 01140-070, Brazil}
\affiliation{Instituto de F\'{i}sica Te\'{o}rica - Universidade Estadual Paulista, S\~{a}o Paulo, SP, 01140-070, Brazil}

\author{Manfred Sigrist}
\affiliation{Institute for Theoretical Physics, ETH Zurich, CH-8093, Zurich, Switzerland}

\begin{abstract}

The character of the superconducting phase of \SRO\, is topic of a longstanding discussion. The classification of the symmetry allowed order parameters has relied on the tetragonal symmetry of the lattice and on cylindrical Fermi surfaces, usually taken to be featureless, not including the non-trivial symmetry aspects related to their orbital content. Here we show how the careful account of the orbital degree of freedom and a three dimensional description lead to a much richer classification of order parameters. We analyse the stability and degeneracy of these new order parameters from the perspective of the concept of \emph{superconducting fitness} and propose new order parameter candidates which can systematically account for the observed phenomenology in this material.
\end{abstract}

\date{\today}

\maketitle
\section{Introduction}
\SRO\, is among the materials with the highest quality single-crystals \cite{MacRMP03,MaoMRB00} and with the best characterized normal state Fermi surfaces \cite{MacPRL96,BerPRL00,MaeJPSJ97,DamPRL00}. Yet, the nature of the superconducting state in this material remains controversial for more than 20 years \cite{MacNJP17}. Experimental evidence from different probes give us conflicting information if we try to understand the phenomenology of this material from the perspective of an order parameter on a single cylindrical Fermi surface. The solution to this conundrum might rely on the fact that \SRO\, is a complex material, since the faithful description of its normal state electronic structure requires at least the three Ru 4d orbitals in the $t_{2g}$ manifold. In contrast to the microscopic description in the orbital basis, superconductivity is usually understood as an instability out of a Fermi surface. When studying superconductivity in \SRO\, it is usual to erase the microscopic complexity needed for the realistic representation of its three Fermi surfaces, and to start treating these as featureless entities \cite{NomJPSJ00,NomJPSJ02,MiyPRL99,KuwPRL00}.

Several experiments have indicated that the order parameter is in the spin-triplet sector, in particular Knight shift \cite{IshNat98,IshPRB15} and neutron scattering measurements \cite{DufPRL00}, which observed no change in the spin susceptibility across the superconducting critical temperature, $T_c$, for any magnetic field direction. Another important piece of evidence is the observation of the onset of time-reversal symmetry breaking (TRSB) at $T_c$ from muon spin rotation \cite{LukNat98,LukPB00} and polar Kerr effect measurements \cite{XiaPRL06}. These two facts together point towards a chiral order parameter with a d-vector $\textbf{d}(\bk) =  (0,0,k_x \pm i k_y)$ \cite{RicJPCM1995, MacRMP03,MaeJPSJ12,KalRPP12}, the only unitary odd-parity triplet order parameter in a tetragonal material to break time-reversal symmetry. Contradictions emerge once we consider complementary experimental results. For example, specific heat \cite{NisJLTP99,NisJPSJ00,DegPRL04,DegJPSJ04,Kit18} and ultrasound \cite{LupPRL01} measurements suggest the presence of horizontal line nodes in the superconducting gap, and new thermal conductivity measurements \cite{HasPRB2017} give evidence for vertical line nodes. In addition, recent experiments are now challenging what were thought to be well stablished results. In particular, novel Knight shift measurements indicate a drop in the spin susceptibility for in-plane magnetic fields, challenging the proposal of an order parameter with a d-vector along the z-direction \cite{Pus19}. Also, latest uniaxial strain experiments performed at the micron-scale observe no obvious splitting of the critical temperature as a function of strain, expected if the order parameter has two-components \cite{WatPRB18}. These recent results motivate us to look more carefully into the possible order parameters for \SRO\, from a microscopic perspective.

This paper is organized as follows: In Section \ref{Sec:H0} we introduce the most general model for the normal state of \SRO\, based on the presence of time-reversal symmetry, the point group symmetry $D_{4h}$ (which includes inversion symmetry) and the nature of the underlying orbital degrees of freedom (DOF) in the $t_{2g}$ manifold. In Section \ref{Sec:D} we reclassify the order parameters, considering explicitly the orbital dependence, discuss their properties and probe these  against the most recent experimental results. In Section \ref{Sec:Fit} we focus on intra-orbital order parameters and apply the concept of \emph{superconducting fitness}, which allows for a qualitatively understanding of the stability of each order parameter. From this analysis, we also discuss the presence of symmetry protected and accidental or quasi-degeneracies and the consequences for experiments under strain. We conclude with Section \ref{Sec:Con}, summarizing our results and highlighting new directions for theoretical investigation for a more complete understanding of \SRO.

\section{The normal state Hamiltonian}\label{Sec:H0}

\SRO\, has the tetragonal space group $I4/mmm$, or $\# 139$ \cite{MacRMP03}. This group consists of operations in the point group $D_{4h}$ and intra-unit-cell shifts by half a lattice parameter in all directions.  Focusing on the point group, here we refer to $D_4$ since in this case the tables below have a more compact form (the product with inversion essentially splits the representations in even and odd). 
It is well known that the important DOF for the description of \SRO\, are the electrons in the $t_{2g}$ orbital manifold  in the Ru ions, namely $d_{yz}$, $d_{xz}$ and $d_{xy}$. Choosing the basis $\Phi^\dagger_\bk  = (c_{yz\uparrow}^\dagger, c_{yz\downarrow}^\dagger,  c_{xz\uparrow}^\dagger, c_{xz\downarrow}^\dagger, c_{xy\uparrow}^\dagger, c_{xy\downarrow}^\dagger)_\bk$,  one can construct the most general three-orbital single-particle Hamiltonian describing the normal state as:
\begin{eqnarray}
\mathcal{H}_0(\bk)=\Phi^\dagger_\bk H_0(\bk) \Phi_\bk,
\end{eqnarray}
with
\begin{eqnarray}\label{Eq:H0}
H_0(\bk)=\sum_{a,b}  h_{ab}(\bk) [\lambda_a \otimes \sigma_b ],
\end{eqnarray}
where $h_{ab}(\bk)$ are 36 real functions of momenta $\bk$, $\lambda_{a=1,...,8}$ are the Gell-Mann matrices and $\lambda_0 = \sqrt{\frac{2}{3}} I_3$, with $I_3$ the three-dimensional identity matrix, standing for the orbital DOF, and $\sigma_{b=1,2,3}$  are Pauli matrices, with $\sigma_0$ the two-dimensional identity matrix, standing for the spin DOF. The explicit form of the Gell-Mann and Pauli matrices used in this work are given in Appendix \ref{Sec:GMM}.

Requiring the Hamiltonian to be invariant under inversion and time-reversal, we find restrictions on the allowed pairs of indices $(a,b)$. Inversion, $P=\sqrt{\frac{3}{2}}\lambda_0\otimes \sigma_0$,  acts trivially on the spin and orbital DOF:
\begin{eqnarray}
P H_0(\bk) P^{-1}=H_0(\bk)
\end{eqnarray}
implying that the functions $h_{ab}(\bk)=h_{ab}(-\bk)$ must all be even in momentum. Time-reversal, $\Theta = K \sqrt{\frac{3}{2}}\lambda_0\otimes (i \sigma_2) $, with $K$ standing for complex conjugation, acts on the spin DOF:
\begin{eqnarray}
\Theta H_0(\bk) \Theta^{-1}=H_0(\bk),
\end{eqnarray}
implying $h_{ab}(\bk) = \pm h^*_{ab}(-\bk)$, with the plus (minus) sign for imaginary (real) products $[\lambda_a \otimes \sigma_b ]$. The explicit pairs of indices $(a,b)$ are summarized  in Table \ref{Tab:ab}. Given that the functions $h_{ab}(\bk)$ are real and must be even in $\bk$, only a subset of 15 pairs $(a,b)$ is in fact allowed in the Hamiltonian.

The Hamiltonian should also be invariant under the point group operations associated with $D_4$. Selecting as generators $C_4$ (rotation along the z-axis by $\pi/2$), $C'_{2}$ (rotation along the x-axis by $\pi$) and $C''_2$ (rotation along the diagonal $x=y$ by $\pi$), the basis matrices $[\lambda_a \otimes \sigma_b ]$ transform as specific irreducible representations of the point group. The invariance of the Hamiltonian $H_0(\bk)$ under the point group requires that the basis functions $h_{ab}(\bk)$ transform under the same irreducible representation as the associated basis matrices. The explicit form of the point group operators are given in Appendix \ref{Sec:D4}, and the result of the symmetry analysis for the normal state Hamiltonian is summarized in Table \ref{Tab:H0}.

\begin{table}[t]
\begin{center}
    \begin{tabular}{| c | c | c |}
    \hline
    $h_{ab}(\bk)$ &   $(a,b)$ & $N$\\ \hline
    Even & $(\{0,1,2,3,7,8\},2)$ and $(\{4,5,6\},\{0,1,3\})$ & 15 \\ \hline
    Odd &  $(\{4,5,6\},2)$ and $(\{0,1,2,3,7,8\},\{0,1,3\})$ & 21  \\ \hline
    \end{tabular}
\end{center}
\caption{Parity of $h_{ab}(\bk)$ functions according to time-reversal symmetry. $N$ indicates the number of pairs $(a,b)$ for each parity.}
\label{Tab:ab}
\end{table}

\begin{table}[t]
\begin{center}
    \begin{tabular}{| c | c | c | c | c |}
    \hline
     Irrep   & $(a,b)$    & Type & Only in 3D \\ \hline
      \multirow{4}{*}{$A_1$}& $(0,0)$ &     intra-orbital hopping & \\ \cline{2-4}    
    & $(8,0)$ &    intra-orbital hopping & \\ \cline{2-4}    
    & $(4,3)$ &    atomic SOC &  \\ \cline{2-4}       
    & $(5,2)-(6,1)$ &  atomic-SOC &   \\ \hline
     $A_2$ & $(5,1)+(6,2)$ &    $\bk$-SOC & $\checkmark$ \\ \hline
     \multirow{2}{*}{$B_1$}& $(7,0)$ &    intra-orbital hopping &   \\  \cline{2-4}    
    & $(5,2)+(6,1)$ &  $\bk$-SOC  & $\checkmark$ \\ \hline
     \multirow{2}{*}{$B_2$}& $(1,0)$ &   inter-orbital hopping &  \\ \cline{2-4}    
    & $(5,1)-(6,2)$ &    $\bk$-SOC & $\checkmark$ \\ \hline
    \multirow{2}{*}{$E(i)$}& $(3,0)$ &   inter-orbital hopping & $\checkmark$  \\ \cline{2-4} 
    & $-(2,0)$ & inter-orbital hopping & $\checkmark$  \\ \hline
    \multirow{2}{*}{$E(ii)$}& $(4,2)$ &    $\bk$-SOC& $\checkmark$ \\ \cline{2-4} 
    & $-(4,1)$ &   $\bk$-SOC& $\checkmark$ \\ \hline
        \multirow{2}{*}{$E(iii)$}&$(5,3)$ &    $\bk$-SOC& $\checkmark$  \\  \cline{2-4} 
    & $(6,3)$ &   $\bk$-SOC& $\checkmark$  \\ \hline
    \end{tabular}
\end{center}
\caption{List of the 15 symmetry allowed terms in the normal state Hamiltonian $H_0(\bk)$ given by Eq.\ref{Eq:H0}. For each $(a,b)$, the basis function $h_{ab}(\bk)$ should transform according to a specific irreducible representation (Irrep), and can be associated with different physical processes (Type). Here $\bk$-SOC stands for even-momentum spin-orbit coupling.  The last column highlights which symmetry allowed terms are present only in three-dimensional models. For the two-dimensional Irreps $E(\alpha)$, $\alpha = i, ii, iii$, the entries are organized such that the first $(a,b)$ term transforms as $x$ and the second as $y$.}
\label{Tab:H0}
\end{table}

Note that the form we find through this symmetry analysis is in accordance with the well stablished Hamiltonian for \SRO\,  \cite{ScaPRB2014, RamPRB16}, in which the terms $(0,0)$, $(7,0)$ and $(8,0)$ are associated with intra-orbital hopping in the $A_1$, $B_1$ and $A_1$ representations, respectively; $(1,0)$ is associated with inter-orbital hopping in $B_2$, allowed only between $xz$ and $yz$ orbitals; and $(4,3)$ and $(5,2)-(6,1)$ in $A_1$ are associated with atomic spin-orbit coupling (SOC). Other allowed terms are: $\{(3,0),(-2,0)\}$ in $E$, related to out-of-plane inter-orbital hopping between $xz$ or $yz$ and $xy$ orbitals; $\{(4,2),-(4,1)\}$ and $\{(5,3),(6,3)\}$ also in $E$, as well $(5,2)+(6,1)$ and $(5,1)\pm(6,2)$ in $B_1$, $A_2$ and $B_2$, respectively, all related to even $k$-dependent SOC, which are usually not taken into account within two-dimensional models. 

\section{The order parameters in the orbital basis}\label{Sec:D}


In multiorbital superconducting systems, the effective Bogoliubov-de Gennes Hamiltonian can be written as:
\begin{eqnarray}\label{Eq:MFH}
H_{BdG}&=&\sum_\bk \Psi_\bk^\dagger \begin{pmatrix}
 H_{0}(\bk)& \Delta(\bk)\\
\Delta^\dagger(\bk) & - H^*_{0} (-\bk)
\end{pmatrix}
\Psi_\bk,
\end{eqnarray}
in terms of the \emph{multi-orbital Nambu spinors}:
\begin{eqnarray}
\Psi_\bk^\dagger = (\Phi^\dagger_{\bk}, \Phi_{-\bk}^T) ,
\end{eqnarray}
with $\Phi^\dagger_{\bk}$ and $H_0(\bk)$ defined above for the case of \SRO. Here $ \Delta(\bk)$ is the gap matrix which can describe both spin singlet and triplet pairing in multi-orbital space.

Similarly to the parametrization of the normal state Hamiltonian, we start parametrizing the gap matrix with 36 functions $d_{ab} (\bk)$:
\begin{eqnarray}
\Delta (\bk) = \sum_{a,b}  d_{ab} (\bk) [\lambda_a \otimes \sigma_b (i \sigma_2)].
\end{eqnarray}
In analogy to the \emph{d-vector} parametrizing the triplet order parameter for a single band superconductor, here we introduce a \emph{d-tensor} notation, $d_{ab} (\bk)$. In order to satisfy the anti-symmetry of the pair wave-function, the order parameter should follow
\begin{eqnarray}
\Delta(\bk) = - \Delta^T(-\bk),
\end{eqnarray}
such that we can separate the functions $d_{ab} (\bk)$ in even- or odd-parity if the tensor product $[\lambda_a \otimes \sigma_b (i\sigma_2)]$ is anti-symmetric or symmetric, respectively. Interestingly the pairs of $[a,b]$ associated with even or odd $d_{ab}(\bk)$ functions are the same as the ones identified for the normal state Hamiltonian parameters $h_{ab}(\bk)$ summarized in Table \ref{Tab:ab}. This can be understood by the fact that the basis matrices chosen here are all Hermitian, such that the transpose is equivalent to the complex conjugate, making the correspondence to time-reversal symmetry. Note that, in order to distinguish the parametrization of the gap matrix from the parametrization of the normal state Hamiltonian, we use different brakets $[a,b]$ for the gap function indexes.

We can further classify the order parameters considering the point group transformations, which rotate the order parameter as $U \Delta(\bk) (U^{-1})^*$ \cite{RamPRB16,RamPRB18}. We start the analysis looking at how the product $[\lambda_a\otimes \sigma_b (i\sigma_2)]$ transforms under the generators of the point group, what allows us to associate these matrices with distinct irreducible representations (Irreps) of $D_4$.  The properties of all basis matrices are listed in detail in Appendix \ref{Sec:OP}. Here we focus on intra-orbital pairing, $a=\{0,7,8\}$, and summarize the properties of the basis matrices in the first four columns of Table \ref{Tab1} . 

\begin{center}
\begin{figure}[t]
\includegraphics[width=0.75\linewidth, keepaspectratio]{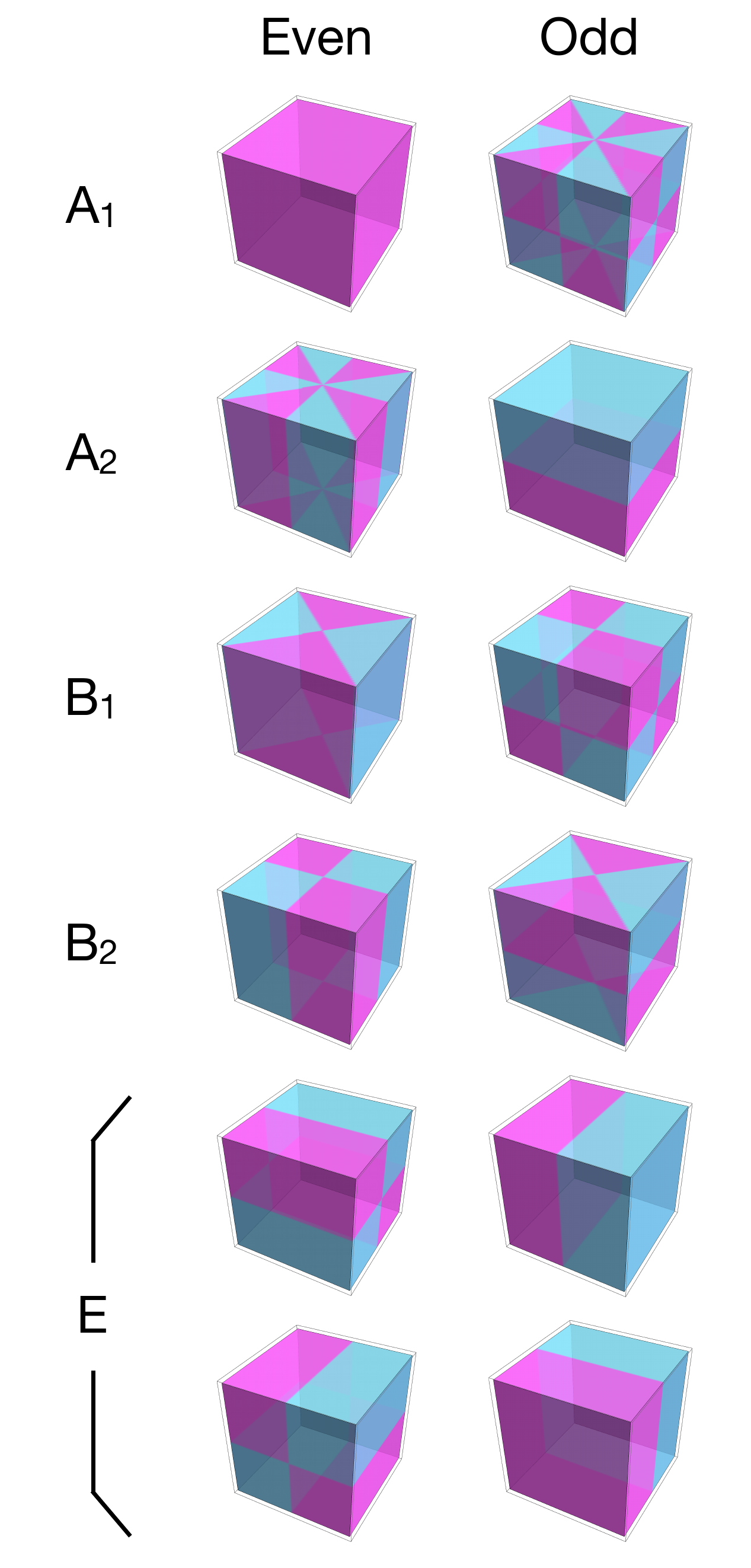}
\caption{Form factors associated with the (lowest power) even and odd basis functions of each of the irreducible representations of the $D_{4}$ point group. The different colors indicate positive and negative values.}
\label{Fig:FF}
\end{figure}
\end{center}

We can now  introduce the non-trivial momentum dependence of $d_{ab}(\bk)$ (see form factors in Fig. \ref{Fig:FF}). In order to determine the Irrep of the complete order parameter, we need to take the product of the Irrep of $d_{ab}(\bk)$ with the Irrep of  $[\lambda_a\otimes\sigma_b (i\sigma_2)]$. The resulting Irrep can be inferred from the character table of the point group (see Appendix \ref{Sec:D4}), and the results are summarized in the last five columns of Table \ref{Tab1} for the intra-orbital components of the order parameter.

\begin{table}[t]
\begin{center}
    \begin{tabular}{c }
    \includegraphics[width=0.48\textwidth]{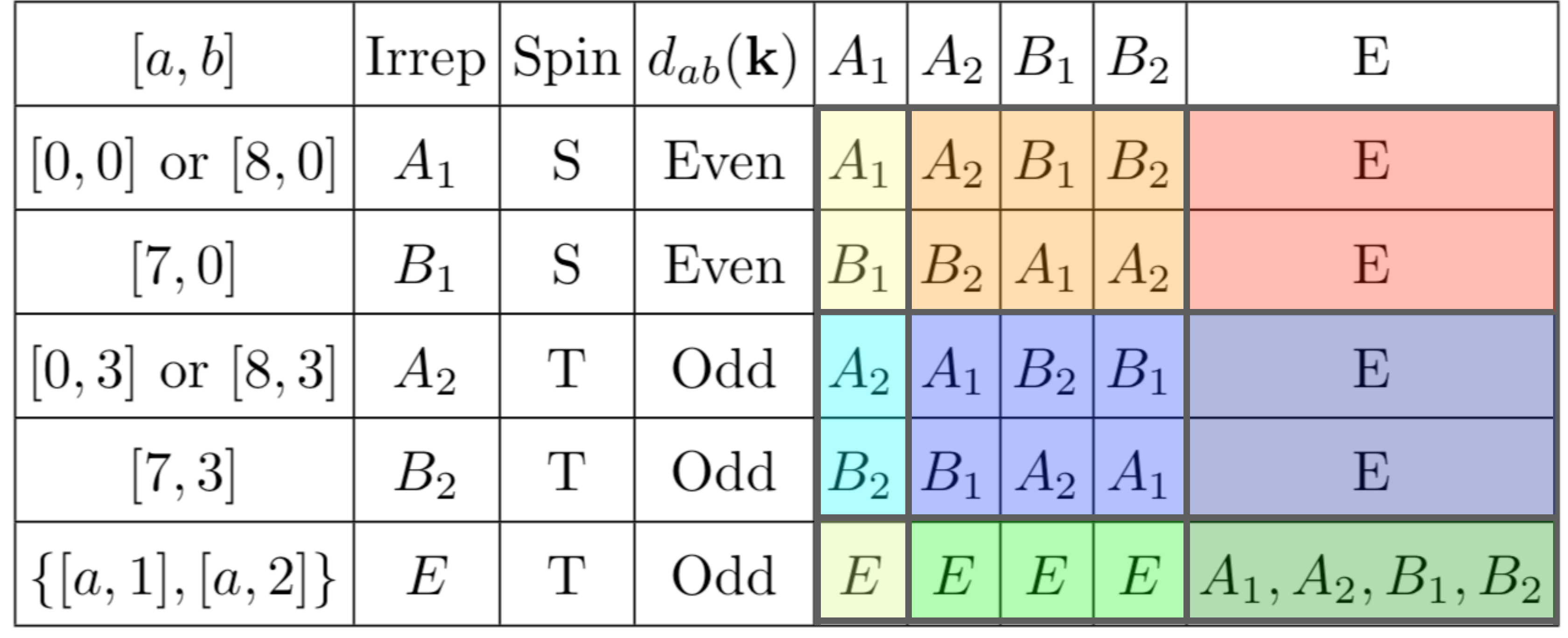}
    \end{tabular}
\end{center}
\caption{Symmetry of the order parameter with matrix basis $[\lambda_a\otimes \sigma_b (i\sigma_2)]$, represented by $[a,b]$ (first column) and function $d_{ab}(\bk)$ in different Irreps (first line of last five columns), for intra-orbital pairing $a=\{0,7,8\}$. The second column gives the representation of the matrix basis, the third column the spin character singlet (S) or triplet (T), and the fourth column the parity of the function $d_{ab}(\bk)$.}
\label{Tab1}
\end{table} 

\subsection{Properties of the order parameters}
We now go over Table \ref{Tab1}, analysing in detail the symmetry properties of the order parameters in different sectors (indicated by different colors), and summarize
the key experimental signatures in Table \ref{TableExp}. In the next section we discuss which of these sectors consistently account for the features observed in the most recent experiments. 

The order parameters in the yellow sector $(i)$ are spin singlets, have even $d_{ab}(\bk)\sim \text{cte}$, do not carry symmetry protected nodes, and are associated with one-dimensional representations. The superconducting states in the orange sector $(ii)$ are also spin singlets, but now $d_{ab}(\bk)\sim k_xk_y(k_x^2-k_y^2)$, $(k_x^2-k_y^2)$ or $k_xk_y$ for even functions in the $A_2$, $B_1$ or $B_2$ representations (according to the first line of Table \ref{Tab1}). All these order parameters are associated with symmetry protected vertical line nodes and once we take the product of the representations of the matrix part $[\lambda_a\otimes\sigma_b (i\sigma_2)]$ with the function $d_{ab}(\bk)$, we find order parameters belonging to one of the one-dimensional representations $A_1$, $A_2$, $B_1$ or $B_2$. The order parameters in the red sector (iii) are also singlets, but now the even basis functions $d_{ab}(\bk)\sim \{k_xk_z, k_yk_z\}$ belong to a two-dimensional representation. Note that now we are guaranteed to have at least horizontal line nodes (assuming that a chiral state would be energetically more favourable, in which the two components appear in a complex superposition eliminating the vertical line nodes). The fact that the matrix part $[\lambda_a\otimes\sigma_b (i\sigma_2)]$ are associated with one-dimensional representations and the function $d_{ab}(\bk)$ is associated with the two-dimensional representation, makes the complete order parameters transform as the two-dimensional representation $E$. 

Moving on now to the triplet states, these are usually parametrized in terms of a d-vector, which gives the direction perpendicular to the orientation of the Cooper pair spin. Here the direction of the d-vector in the standard notation is encoded in the label $b$ in $[a,b]$, as can be inferred from the matrix form $[\lambda_a\otimes\sigma_b (i\sigma_2)]$. In the cyan sector $(iv)$ we find triplet states with the d-vector along the z-direction, corresponding to in-plane equal spin pairing. In this configuration, an in-plane magnetic field does not lead to paramagnetic depairing, in contrast to a field along the z-axis. In this sector, the odd function $d_{ab}(\bk)\sim k_xk_yk_z(k_x^2-k_y^2)$ encodes both horizontal and vertical line nodes, and the order parameters fall in the one-dimensional representations $A_2$ or $B_2$. In the blue sector $(v)$ we also find triplet states with a d-vector along the z-direction, but now the odd form factor $d_{ab}(\bk)\sim k_z$, $k_zk_xk_y$ or $k_z(k_x^2-k_y^2)$ in the $A_2$, $B_1$ or $B_2$ representations, respectively, with only horizontal or both horizontal and vertical symmetry protected line nodes. The order parameters in this sector fall in one of the one-dimensional representations $A_1$, $A_2$, $B_1$ or $B_2$. The dark blue sector $(vi)$ includes triplet order parameters with a d-vector along the z-direction, and come with odd basis functions $d_{ab}(\bk)\sim \{k_x, k_y\}$ associated with the two-dimensional representation $E$. For this sector the presence of nodes is not guaranteed since a chiral state is expected to be energetically more favourable, and the complete order parameter belongs to the two-dimensional representation $E$.

\begin{table}[t]
\begin{center}
    \begin{tabular}{c }
    \includegraphics[width=0.25\textwidth]{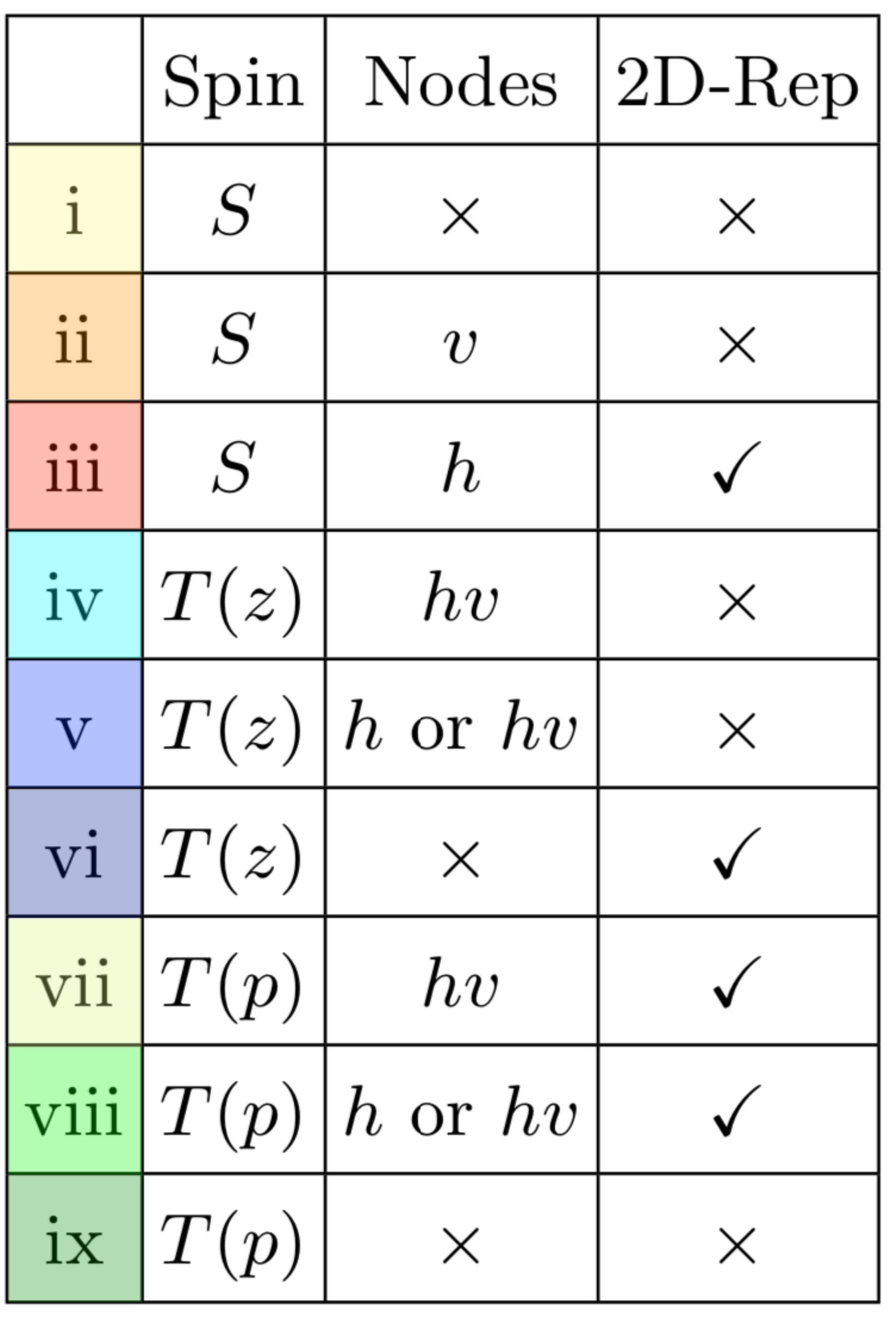}
    \end{tabular}
\end{center}
 \caption{Summary of the key experimental signatures for the different families of superconducting order parameters listed in Table \ref{Tab1}. (i-ix) can be identified by the color scheme or by reading the sectors highlighted by thick lines from left to right, top to bottom in Table \ref{Tab1}. Here $S$ stands for singlet and $T$ for triplet (z and p correspond to the d-vector direction along the z-axis or in plane), $h$ stands for horizontal, $v$ for vertical and $hv$ for simultaneous $h$ and $v$ line nodes.}
\label{TableExp}
\end{table} 

The last row in Table \ref{Tab1} includes triplet order parameters with a d-vector in plane. For this configuration, a magnetic field along the z-direction does not cause paramagnetic limiting, while an in-plane field breaks the pairs. Note also that now the matrix part of the order parameter forms itself a two-dimensional representation, denoted by $\{[a,1], [a,2]\}$ in the table, indicating that the matrices $[a,1]$ and $[a,2]$ transform as basis of the two-dimensional representation $E$ under the point group operations. In the light green sector $(vii)$ these basis matrixes are combined with an odd form factor $d_{ab}(\bk)\sim k_xk_yk_z(k_x^2-k_y^2)$ encoding both horizontal and vertical line nodes. The product of the matrix component of the order parameters belonging to the two-dimensional representation $E$ with a form factor $d_{ab}(\bk)$ in a one-dimensional representation leads to an order parameter belonging to the two-dimensional representation $E$. The order parameters in the green sector $(viii)$ are also triplet states with an in-plane d-vector, but now the odd form factors $d_{ab}(\bk)\sim k_z$, $k_zk_xk_y$ or $k_z(k_x^2-k_y^2)$ in the $A_2$, $B_1$ or $B_2$ representations, respectively, have only horizontal or both horizontal and vertical symmetry protected line nodes. Given the product of the representations of the matrix and the form factor components of the order parameters, the complete states in this sector transform within the two-dimensional representation $E$. Finally, the order parameters in the dark green sector $(ix)$ are also triplet states with an in-plane d-vector, now with odd basis functions $d_{ab}(\bk)\sim \{k_x, k_y\}$ associated with the two dimensional representation $E$. Given that now both the matrix content and the form factors belong to the two-dimensional representation $E$, we need to consider the product $E\times E$ to define the Irrep of the complete order parameter. This product is decomposed in the one-dimensional Irreps $A_1$, $A_2$, $B_1$ or $B_2$, as indicated in Table \ref{Tab1}. Note that, in analogy to the helical states in the original classification, these states are also fully gapped.


\subsection{Connection with recent experiments}\label{Sec:OPExp}

The most recent Knight shift measurements indicate that there is a substantial drop in the spin susceptibility across the superconducting transition for in-plane magnetic fields \cite{Pus19}. This observation is not easy to reconcile with a triplet state with a d-vector along the z-direction, whose in-plane spin susceptibility is not expected to change as the superconducting state sets in. Based on this fact, order parameters in the blue sectors $(iv)$, $(v)$ and $(vi)$, which include the originally proposed chiral p-wave state, seem not to be good candidates. Considering now the evidence for line nodes from specific heat \cite{NisJLTP99,NisJPSJ00,DegPRL04,DegJPSJ04,Kit18}, ultrasound attenuation \cite{LupPRL01} and recent thermal transport \cite{HasPRB2017}, gap structures without symmetry protected nodes in the $(i)$ and $(ix)$ sectors do not seem to satisfy the constraints imposed by the observations. Furthermore, muon spin rotation \cite{LukNat98,LukPB00} and polar Kerr effect \cite{XiaPRL06} experiments observe the onset of time-reversal symmetry breaking, and ultrasound measurements \cite{LupPRL01} observe a drop in the shear modulus below the critical temperature. These two observations, from experiments of very different nature, in principle require an order parameter belonging to a two-dimensional Irrep \cite{Sigrist1991}.

After these considerations, the order parameters which are in accordance with the observed phenomenology belong to sectors $(iii)$, $(vii)$ or $(viii)$. Note that these always carry horizontal line nodes (some also with vertical line nodes). We can now write explicitly the form of the intra-orbital components of the best candidate order parameters. Starting with the singlet states in sector $(iii)$, we have a TRSB state with horizontal line nodes:
\begin{eqnarray}\label{DSROS}
\Delta_{SRO}^S (\bk) &=&  \sum_a   \left(d_{a}^{xz}(\bk)+\alpha d_{a}^{yz}(\bk) \right)\lambda_a\otimes \sigma_0(i\sigma_2),
\end{eqnarray}
where $a=\{0,7,8\}$, $\alpha$ is a complex coefficient and the functions $d_{a}^{X}(\bk)$ transform as $X$ under the point group operations. For the triplet states in sectors $(vii)$ or $(viii)$, we write TRSB triplet states with in-plane d-vector and horizontal and possibly also vertical line nodes:
\begin{eqnarray}\label{DSROT}
\Delta_{SRO}^T (\bk) &=&  \sum_a   d_{a}^X(\bk) \left(\lambda_a\otimes \sigma_1+ \alpha  \lambda_a\otimes \sigma_2 \right) (i\sigma_2),
\end{eqnarray}
where $X=\{A_1,A_2,B_1,B_2\}$.


\section{Superconducting Fitness analysis}\label{Sec:Fit}
The concept of superconducting fitness proved itself useful for the understanding of the effects of external symmetry breaking fields in complex multi-orbital superconductors, and also gives a measure of the intrinsic robustness of different superconducting states within a given normal state electronic structure \cite{RamPRB16, Ramires_2017, RamPRB18}. The superconducting fitness functions are defined as \cite{RamPRB16,RamPRB18}:
\begin{eqnarray}
F_C(\bk) &=& H_0(\bk) \Delta (\bk) - \Delta(\bk) H_0^*(-\bk),\\ \nonumber
F_A(\bk) &=& H_0(\bk) \Delta (\bk) + \Delta(\bk) H_0^*(-\bk).
\end{eqnarray}
Note that these are in fact matrices, which take into account the normal state electronic structure in $H_0(\bk)$ and the superconducting gap in $\Delta (\bk)$, both encoding as many microscopic degrees of freedom as needed. The averages over the Fermi surface of $Tr[|F_A(\bk)|^2]$ and $Tr[|F_C(\bk)|^2]$ were shown to directly determine the critical temperature for two-orbital scenarios \cite{RamPRB18}. The larger $F_A(\bk)$, the higher the critical temperature, while a finite $F_C(\bk)$ introduces detrimental effects to the superconducting state, reducing the critical temperature. 

We apply this framework to \SRO, and the results for $F_A(\bk)$ and $F_C(\bk)$ are summarized in Table \ref{TabFit} and \ref{TabFitC}, respectively. We highlight that, among the intra-orbital order parameters, the terms which contribute to a finite $F_C(\bk)$ in the standard two-dimensional model are: $(1,0)$, associated with inter-orbital hopping, carrying a form factor in $B_2$ (even); and $(4,3)$ and $(5,2)-(6,1)$ in $A_1$ (even), associated with atomic SOC. In order to reduce the detrimental effects introduced by a finite $F_C(\bk)$ function, we would like to combine these terms with order parameters with non-trivial form factors $d_{ab}(\bk)$, preferably with nodal basis functions orthogonal to $B_2$ (even), for both singlet and triplet states.
Analysing now $F_A(\bk)$, we focus on the largest contributions to the normal state Hamiltonian, given by the intra-orbital hopping terms $(7,0)$ in $B_1$ (even) and $(8,0)$ in $A_1$ (even). In order to maximize the average of $Tr[|F_A(\bk)|^2]$ over the Fermi surface for the singlet states, order parameters with $d_{ab}(\bk)$ in $A_1$ (even) would be preferred, but under the condition that these should be nodal in order to minimize $Tr[|F_C(\bk)|^2]$, the best form factor would be in the $B_1\sim (k_x^2-k_y^2)$ (even) channel. For triplet states, to maximize $Tr[|F_A(\bk)|^2]$, a form factor in $B_2\sim (k_x^2-k_y^2)k_z$ (odd) is the most favoured. Concerning the choice of singlet versus triplet states, the superconducting fitness analysis finds singlet states to be the most robust, what is guaranteed by atomic SOC, as can be inferred by the larger coefficient for a-SOC in the first line of Table \ref{TabFit}.

\begin{table}[t]
\begin{center}
    \begin{tabular}{| c | c | c | c | c | c | c | c | c | c | c | }
    \cline{1-8}
    $Tr[|F_A(\bk)|^2]$ & \multicolumn{5}{c|}{2D} & \multicolumn{2}{c|}{3D}\\ \hline
    $[a,b]$ & (1,0) & (7,0) & (8,0) & a-SOC & 2D-deg  & IOH-z & k-SOC \\ \hline
    $[0,0]$ &  1 & 1 & 1 & 3 & &  2 & 4   \\ \hline
    $[7,0]$ &  - & 3/2 & 1/2 & 3/4  &  &  3/4 & 3/4    \\ \hline
    $[8,0]$ &  1/2 & 1/2 & 3/2 &  3/4 & &   1/4 & 5/4     \\ \hline
    $[0,3]$ &  1 & 1 & 1 &  1& * & 2 & 2   \\ \hline
    $[7,3]$ &  - & 3/2 & 1/2 &  3/4 & (**) &  3/4 & 15/4     \\ \hline
    $[8,3]$ &  1/2 & 1/2 & 3/2 &  11/4 & (***) &  1/4 & 1/4     \\ \hline 
    $\{[0,1],[0,2]\}$ &  1 &  1 & 1 & 1 & * &  2 & 1   \\ \hline
    $\{[7,1],[7,2]\}$ &  - &  3/2 & 1/2 & 9/4 & (**) &  3/4 & 9/4     \\ \hline
    $\{[8,1],[8,2]\}$ &  1/2 &  1/2 & 3/2 & 5/4 & (***) &  1/4 & 11/4      \\ \hline     
    \end{tabular}
\end{center}
  \caption{Superconducting Fitness analysis for different matrix basis $[a,b]$ of $\Delta(\bk)$ indicated in the first column. The table displays the contribution of each term $(c,d)$ in $H_0(\bk)$ for the superconducting fitness parameters as $Tr[|F_A(\bk)|^2] = \frac{64}{9} \sum_{cd} (\text{table entry})|d_{ab}(\bk)|^2|h_{cd}(\bk)|^2$. Columns 2-5 include terms present in a two dimensional effective model, while columns 7-8 introduce additional terms allowed in a three dimensional model. Here a-SOC stands for atomic SOC, associated with terms $(4,3)$ and $(5,2)-(6,1)$; IOH-z stands for inter-orbital inter-plane hopping associated with $\{(3,0),-(2,0)\}$, k-SOC is associated with momentum-dependent SOC from  $\{(4,2),-(4,1)\}$ and $\{(5,3),(6,3)\}$. The column labelled 2D-deg indicates by asterisks which pairs of order parameters are degenerate for a two dimensional model (quasi-degeneracies are indicated by asterisks in parenthesis). Note that for a three dimensional model no degeneracies are left.}
  \label{TabFit}
\end{table}

\begin{table}[t]
\begin{center}
    \begin{tabular}{| c | c | c | c | c | c | c | c | c | c | c | }
    \cline{1-8}
    $Tr[|F_C(\bk)|^2]$ & \multicolumn{5}{c|}{2D} & \multicolumn{2}{c|}{3D}\\ \hline
    $[a,b]$ & (1,0) & (7,0) & (8,0) & a-SOC & 2D-deg  & IOH-z & k-SOC \\ \hline
    $[0,0]$ &  - & - & - & - & &  - & -   \\ \hline
    $[7,0]$ &  3/2 & - & - & 9/4  &  & 3/4  & 15/4    \\ \hline
    $[8,0]$ &  - & - & - &  9/4 & & 9/4   &  9/4   \\ \hline
    $[0,3]$ &  - & - & - &  2& * & - &  2  \\ \hline
    $[7,3]$ &  3/2 & - & - &  9/4 & (**) & 3/4  &  3/4     \\ \hline
    $[8,3]$ &  - & - & - &  1/4 & (***) & 9/4  &   13/4   \\ \hline 
    $\{[0,1],[0,2]\}$ &  - &  - & - & 2 & * &  - & 3   \\ \hline
    $\{[7,1],[7,2]\}$ & 3/2 &  - & - & 9/4 & (**) & 3/4   & 9/4     \\ \hline
    $\{[8,1],[8,2]\}$ &  - &  - & - & 7/4 & (***) & 9/4   &  3/4    \\ \hline     
    \end{tabular}
\end{center}
  \caption{Superconducting Fitness analysis. Contributions for the quantity $Tr[|F_C(\bk)|^2]$. Same notation as Table \ref{TabFit}.}
  \label{TabFitC}
\end{table}

\subsection{Order parameter degeneracy}

From Table \ref{TabFit} we can also review the discussion about the accidental degeneracies of the order parameters. We start considering a two dimensional model. It is usually stated that the helical and chiral order parameters are degenerate up to the inclusion of SOC. This argument can be based on a phenomenological Ginzburg-Landau theory, with SOC being introduced at the free energy level by evaluating the expectation value of $\boldsymbol{L}\cdot \boldsymbol{S}$ for a given pair wave-function \cite{Sigrist96,Sigrist05}, or from the analysis of its effects on different pairing mechanisms \cite{YanJPSJ03, AnnPRB06, NgEPL00}. Here we analyze the degeneracy directly from a microscopic perspective, considering an orbital and spin symmetric microscopic interaction. According to the concept of superconducting fitness \cite{RamPRB18}, in the weak coupling limit, the critical temperature depends only on the averages over the Fermi surface of the superconducting fitness parameters. In the context of two dimensional models, we find that the order parameters marked with one asterisk in Tables \ref{TabFit} and \ref{TabFitC}  have exactly the same $F_A(\bk)$ and $F_C(\bk)$ (assuming the same form factors $d_{ab}(\bk)$). This means that these order parameters are in fact degenerate and would have the same critical temperature within a single band scenario (or with three equivalent bands) . This perspective tells us that  atomic SOC does not split the degeneracy between different d-vector directions, and suggests that a rotation of the d-vector in presence of magnetic field is possible. In addition, we would like to highlight that the contributions of inter-orbital hopping $\sim 0.01 t$ and SOC $\sim 0.1t$ introduce shifts of $10^{-2}-10^{-4}$ to $Tr[|F_A(\bk)|^2]$, taking as baseline intra-orbital hopping terms (here $t$ stands for the maximal intra-layer intra-orbital hopping amplitude \cite{ScaPRB2014}). If we neglect these shifts, we have quasi-degeneracies which are not protected by symmetry, therefore not usually discussed (marked by asterisks between parenthesis in Tables \ref{TabFit} and \ref{TabFitC}). This fact is related to the almost degenerate superconducting states recently found in numerical approaches \cite{ScaPRB2014}.  There is also an apparent unexpected degeneracy between singlet and triplet order parameters $[3,0]$ and $[3,3]$. A degeneracy would assume the same form factor $d_{ab}(\bk)$, what is not possible given the different parity of the order parameter in both cases. 

\subsection{Order parameters coupling to the lattice}

It is interesting now to consider some consequences of these symmetry considerations and quasi-degeneracies for the interpretation of experiments under uniaxial strain $\sim \epsilon_{xx}-\epsilon_{yy}$, in which case the point group is reduced from $D_4$ to $D_2$ \cite{WatPRB18}. Given that inversion and time-reversal symmetries are preserved, there are no new $(a,b)$ terms allowed in $H_0(\bk)$, and the matrix basis $[a,b]$ for the order parameters can be separated in even and odd momenta as before. As $D_4$ is reduced to $D_2$, we can make the following correspondence of Irreps: $A_1$ and $B_1 \rightarrow A$, $A_2$ and $B_2\rightarrow B_1$ and $E\rightarrow \{B_2, B_3\}$. The last correspondence means that the two-dimensional representation $E$ of $D_4$ splits into two one-dimensional representations of $D_2$, $B_2$ and $B_3$, under strain.

If we consider the degenerate (and quasi-degenerate) basis matrixes suggested by the superconducting fitness analysis, marked with asterisks in Tables \ref{TabFit} and \ref{TabFitC}, we find that these indicate a (quasi-)degeneracy of order parameters with a d-vector in-plane and along the z-direction for any type of intra-orbital pairing. Consulting Table \ref{Tab1} assuming the same form factor $d_{ab}(\bk)$ for both  order parameters, we find that this implies a (quasi-)degeneracy of order parameters in a one-dimensional representation and in the two-dimensional representation $E$. Under strain, these would necessarily map into different Irreps of $D_2$, corresponding to the splitting of the superconducting transition.

Interestingly, the mapping of Irreps from $D_4$ to $D_2$ suggests a peculiar possibility: if there are quasi-degenerate order parameters in the $A_i$ and $B_i$ representations, $i=\{1,2\}$, the introduction of strain would not lead to a splitting of the superconducting transition since in $D_2$ these belong to the same Irrep. Going beyond the quasi-degeneracies suggested by the superconducting fitness analysis, we can look at more exotic cases of accidental degeneracies. For example, consulting Table \ref{Tab1}, if we choose a form factor $d_{ab}(\bk)$ in $B_2$, its composition with basis matrixes $[0,3]$ or $[8,3]$ in $A_2$, would generate an order parameter in the $B_1$ representation, while its composition with the basis matrix $[7,3]$ in $B_2$ would generate an order parameter in the $ A_1$ representation. Note that all these basis matrices correspond to intra-orbital pairing and d-vector along the z-direction, with the difference that $[0,3]$ and $[8,3]$  introduce intra-orbital order parameters for orbitals $xz$ and $yz$ with same amplitude and phase, while $[7,3]$ introduces intra-orbital order parameters for orbitals $xz$ and $yz$ with same amplitude but opposite phase. Note that if the magnitude of the order parameters for these two orbitals is not the same, the order parameter should necessarily include basis matrices $[a,3]$ ($a=0,8)$  and $[7,3]$, and as a consequence the order paramater would have components both in the $B_1$ and $A_1$ representations. Note that this is in agreement with the mapping of the Irreps from $D_4$ to $D_2$. This scenario is likely to happen around defects and interfaces, where the two orbitals become inequivalent. Given this example, we summarize in Table \ref{Tab:Split} the different types of quasi-degeneracies and their consequences for experiments under strain. 

\begin{table}[t]
\begin{center}
    \begin{tabular}{| c | c | c | c |}
    \hline
    Accidental degeneracy & Product & Under strain & Couples to $c_{66}$\\ \hline
    $A_1$ and $A_2$ & $A_2$ & Split & No  \\ \hline
    $A_1$ and $B_1$ & $B_1$ & No Split & No  \\ \hline
    $A_1$ and $B_2$ & $B_2$ & Split & Yes  \\ \hline
    $A_2$ and $B_1$ & $B_2$ & Split & Yes  \\ \hline
    $A_2$ and $B_2$ & $B_1$ & No Split & No  \\ \hline
    $B_1$ and $B_2$ & $A_2$ & Split & No  \\ \hline
    \{$A_1,A_2,B_1,B_2$\} and $E$ & $E$ & Split & Yes  \\ \hline
    \end{tabular}
\end{center}
\caption{Summary of the experimental consequences of accidental degeneracies under strain, indicating if these split or do not split, as well as if the respective combination of order parameters would allow for coupling to the $c_{66}$ mode in ultrasound attenuation experiments.}
\label{Tab:Split}
\end{table}

Another interesting experiment which allows us to infer about the symmetry of the order parameter as it couples to the lattice is ultrasound attenuation.  Experiments have observed a sharp decrease of the transverse sound velocity, related to the elastic constant $c_{66}$, in $B_2$, across the critical temperature \cite{LupPRL01}. The coupling of the order parameter to a lattice mode with $B_2$ symmetry is in principle expected only if the order parameter is in the two-dimensional representation $E$ \cite{Sigrist1991}. As a more unusual possibility, we can consider the accidental degeneracies discussed above in order to check in which cases a coupling to $c_{66}$ is allowed. The results are summarized in Table \ref{Tab:Split}. If there is an accidental degeneracy between $A_1$ and $B_2$, or between $A_2$ and $B_1$ such coupling would be possible. Note, though, that there is no accidental degeneracy that would not lead to a splitting of the critical temperature under uniaxial strain and at the same time would lead to coupling to the mentioned transverse lattice mode. This result might indicate that further microscopic work concerning the  coupling of different order parameters to the lattice might be required in order to understand if there are new selection rules that emerge due to the multi-orbital nature of the order parameter.

\subsection{Time-reversal symmetry}

Time-reversal symmetry breaking has been observed below $T_c$ by muon spin rotation \cite{LukNat98,LukPB00} and polar Kerr effect measurements \cite{XiaPRL06}. TRSB is associated with multi-component order parameters, and in the case of \SRO\, it has been specially related to the chiral p-wave state in the $E$ representation. Given that the standard chiral p-wave state does not seem to account for several other experimental observations, we now need to consider other two-dimensional order parameters in order to understand this phenomen. As discussed in Section \ref{Sec:OPExp}, good candidates are the singlet chiral d-wave state, or triplet two dimensional representations with in-plane d-vector. Another possibility is the composition of two quasi-degenerate order parameters as discussed in the previous section, what would fit well with the absence of splitting at the transition for certain combinations of Irreps. Recent works have proposed that a singlet order parameter of the type $s+id_{x^2-y^2}$ \cite{Romer2019} would also be a possibility. This type of superposition is beyond what we can infer based on the symmetry analysis of our study.

\section{Discussion and Conclusion}\label{Sec:Con}

From the extended classification of the order parameters for \SRO\, in the orbital basis, we find that the best spin singlet candidate order parameter is given by Eq. \ref{DSROS} with $d_{ab}(\bk)$ in $E \sim \{k_xk_z,k_yk_z\}$, while the best triplet order parameter is the one in Eq. \ref{DSROT} with $d_{ab}(\bk)$ in $B_2 \sim (k_x^2-k_y^2)k_z$. Given the $k_z$ dependence of these order parameters, it is now important to carefully consider inter-layer couplings, within a full three dimensional model. Recent works highlight the non-trivial effects of the third dimension in \SRO, within which one can identify a non-trivial texture in the spin and orbital DOF along the Fermi surface \cite{HavPRL08, VeePRL14, FatPRB15}. Previous theoretical proposals already have suggested order parameters which are odd along the z-direction \cite{ZhiPRL01,AnnPRB02, WenPRL18, RoisingPRB2018}, and therefore should be closely revisited.

As a first step in this direction, we have evaluated the superconducting fitness functions including the out-of-plane terms in the normal state Hamiltonian which are usually neglected. One interesting finding is that, within the assumption of a spin- and orbital-independent interaction, only the inter-layer terms introduce a splitting between triplet states with d-vector along the z-direction or in-plane. Note that this effect is expected to be smaller than the one previously discussed splitting of the triplet order parameters based on the value of the atomic SOC.

Richer possibilities can emerge when we consider the contribution of inter-orbital pairing \cite{HuangArxiv19, GingrasArxiv18}. As can be inferred from the tables in Appendix \ref{Sec:OP}, there is a series of inter-orbital order parameters which fall in one of the five Irreps of the point group, and will therefore coexist with intra-orbital components discussed here. A similar analysis of the superconducting fitness functions can indicate which basis matrices are degenerate for a given Irrep of the form factors $d_{ab}(\bk)$. A construction of a detailed Ginzburg-Landau functional from a microscopic perspective could elucidate what is the most suitable superposition of these different basis matrices. This is an important direction for future work in order to better understand the nature of the order parameter in \SRO\,, and how the different building blocks of the order parameter couple to external fields. 

In conclusion, we analysed \SRO\, from a microscopic perspective, with the most general single-particle Hamiltonian describing the normal state based on the orbitals in the $t_{2g}$ manifold and reclassifying the order parameters in the orbital basis. These constructions use the point group symmetry and on the orbital character of the underlying DOFs. We propose new order parameter candidates which allow for the consistent understanding of many experimental results available at the moment. From the observed phenomenology, the best candidate order parameters are: a singlet state  or a triplet superconductor with an in-plane d-vector. From the superconducting fitness analysis, we determine that the most robust order parameter is a trivial singlet state with form factors in the $B_1$ representation. Among the triplet states,  an order parameter with a form factor in the $B_2$ representation is the most robust, what would imply the presence of  both horizontal and vertical line nodes. Furthermore, we find that for a two-dimensional model with orbital and spin symmetric interactions, the order parameter with in plane d-vector is in fact degenerate with the triplet state with d-vector along the z-direction even in presence of atomic SOC. Interestingly, this degeneracy is lifted only by inter-layer processes. Extra quasi-degeneracies can also be identified, and could be associated with unusual types of order parameter superpositions. Our work does not concern the pairing mechanism, but provides a detailed classification of the order parameters from the orbital perspective and probe these against the available experimental results and within the concept of superconducting fitness. Our analysis should motivate a reconsideration of theories which take into account the role of inter-layer processes and the construction of interacting models from an orbital perspective, considering the role of Hund's coupling \cite{CuocoPRB98, StraArxiv19}, in order to elucidate the origin of superconductivity in \SRO\, from a microscopic perspective.

\acknowledgements
We would like to thank D. Agterberg, W. Chen, E. Hassinger, C. Hicks, K. Ishida, A. Mackenzie and C. Timm for productive discussions. AR acknowledges the support of the grant 2018/18287-8 S\~{a}o Paulo Research Foundation (FAPESP), FUNDUNESP process 2338/2014-CCP. AR is also grateful for the hospitality of the Pauli Centre of ETH Zurich. MS is grateful for financial support by the Swiss National Science Foundation through Division II (No. 184739).

\newpage
\appendix

\section{Gell-Mann Matrices and Pauli Matrices}\label{Sec:GMM}

The Gell-Mann matrices used in this work are the following:

\begin{eqnarray}
\lambda_1 &=& \begin{pmatrix}
0 & 1 & 0 \\
1 & 0 & 0 \\
0 & 0 & 0
\end{pmatrix}, \hspace{1cm}
\lambda_2 = \begin{pmatrix}
0 & 0 & 1 \\
0 & 0 & 0 \\
1 & 0 & 0
\end{pmatrix}, \\ \nonumber
\lambda_3 &=& \begin{pmatrix}
0 & 0 & 0 \\
0 & 0 & 1 \\
0 & 1 & 0
\end{pmatrix}, \hspace{1cm}
\lambda_4 = \begin{pmatrix}
0 & -i & 0 \\
i & 0 & 0 \\
0 & 0 & 0
\end{pmatrix}, \\ \nonumber
\lambda_5 &=& \begin{pmatrix}
0 & 0 & -i \\
0 & 0 & 0 \\
i & 0 & 0
\end{pmatrix}, \hspace{1cm}
\lambda_6 = \begin{pmatrix}
0 & 0 & 0 \\
0 & 0 & -i \\
0 & i & 0
\end{pmatrix}, \\ \nonumber
\lambda_7 &=& \begin{pmatrix}
1 & 0 & 0 \\
0 & -1 & 0 \\
0 & 0 & 0
\end{pmatrix}, \hspace{1cm}
\lambda_8 = \frac{1}{\sqrt{3}}\begin{pmatrix}
1 & 0 & 0 \\
0 & 1 & 0 \\
0 & 0 & -2
\end{pmatrix}. \hspace{1cm}
\end{eqnarray}

We also define
\begin{eqnarray}
\lambda_0 = \sqrt{\frac{2}{3}}\begin{pmatrix}
1 & 0 & 0 \\
0 & 1 & 0 \\
0 & 0 & 1
\end{pmatrix}.
\end{eqnarray}

Note that all matrices follow $Tr[\lambda_j^2] = 2$.

The Pauli matrices used are the following:
\begin{eqnarray}
\sigma_1 &=& \begin{pmatrix}
0& 1  \\
1 & 0 
\end{pmatrix}, \hspace{0.25cm}
\sigma_2 = \begin{pmatrix}
0 & -i \\
i & 0 \\
\end{pmatrix}, \hspace{0.25cm}
\sigma_3 = \begin{pmatrix}
1 & 0 \\
0 & -1 \\
\end{pmatrix},\hspace{0.5cm}
\end{eqnarray}
and
\begin{eqnarray}
\sigma_0 = \begin{pmatrix}
1 & 0 \\
0 & 1 \\
\end{pmatrix}.
\end{eqnarray}

Note that all matrices follow $Tr[\sigma_j^2] = 2$.

\section{Point group $D_4$}\label{Sec:D4}

As stated in the main text, \SRO\, has the tetragonal space group $I4/mmm$, or $\# 139$. This group consists of operations in the point group $D_4$ (total of 8 operations), as well as its combination with inversion (totalling 16 operations). There are also another set of 16 operations which are related to intra-unit-cell shifts by half lattice constant in all directions.  

Given the strong two-dimensional phenomenology of this material, \SRO is usually described by a model on the square lattice, what would suggest $C_4$ symmetry, but here the transformations which consider rotations along in-plane axes are also important because of the odd character of some of the orbitals along the z-direction. We therefore start with $D_4$ symmetry, a group which has 8 elements arranged in five conjugacy classes:
\begin{itemize}

\item E: Identity.

\item 2$C_4$: two rotations along the z-axis, one by $\pi/2$ and another by $3\pi/2$.

\item $C_2$: a rotation along the z-axis by $\pi$.

\item 2$C_2'$: rotations by $\pi$ along the x- or y- axis.

\item 2$C_2''$: rotations by $\pi$ along the diagonals $d(x=y)$ or $\bar{d}(x=-y)$.

\end{itemize}

Given the five conjugacy classes, there are five irreducible representations. Below we have the character table of $D_4$ highlighting the irreducible representations and the associated even and odd lowest order basis functions:
\begin{center}
    \begin{tabular}{| c | c | c | c | c | c | c | c | }
    \hline
    $D_4$ & $E$  &   $2C_4$   &   $C_2$   & $2C_2'$   & $2C_2'' $ & Even Basis & Odd Basis \\ \hline
    $A_1$ &  +1 &  +1  & +1 &  +1 &  +1 & $1$&$ xyz(x^2-y^2)$  \\ \hline
        $A_2$ &  +1 &  +1  & +1 &  -1 &  -1 & $xy(x^2-y^2)$&$ z$   \\ \hline
            $B_1$ &  +1 &  -1  &  +1 &  +1  & -1 &  $x^2-y^2$&$ xyz$ \\ \hline
                $B_2$ &  +1 &  -1  & +1 &  -1  & +1 & $xy$&$ z(x^2-y^2)$ \\ \hline
                    $E$ &  +2 &  0  & -2 &  0 &  0 & $\{xz,yz\}$&$ \{x,y\}$   \\ \hline
    \end{tabular}
\end{center}

Note that all the operations can be written in terms of $C_4, C_{2x}', C_{2d}''$:
\begin{itemize}

\item $E = C_4^4$,

\item $C_2 = C_4^2$,

\item $C_{2y}' = C_4 C_{2x}' C_4^{-1}$.

\item $C_{2\bar{d}}'' = C_4 C_{2d}'' C_4^{-1}$,

\end{itemize}
so if the system is invariant under $C_4, C_{2x}', C_{2d}''$, it is invariant under all transformations of the point group. One can think of these operations as the generators of the group. Note that we should consider also inversion $P$ to complete the point group $D_{4h}$ associated with I4/mmm.

\subsection{Generators acting on coordinates}

The generators identified above act on the spatial coordinates as follows:
\begin{eqnarray}
C_4 = \left\{
                \begin{array}{ll}
                x \rightarrow y\\
                y \rightarrow -x\\
                z \rightarrow z
                \end{array}
              \right.,  \hspace{1cm} 
              C_{2x}' = \left\{
                \begin{array}{ll}
                x \rightarrow x\\
                y \rightarrow -y\\
                z \rightarrow -z
                \end{array}
              \right., 
\end{eqnarray} 
\begin{eqnarray}
              C_{2d}'' = \left\{
                \begin{array}{ll}
                x \rightarrow y\\
                y \rightarrow x\\
                z \rightarrow -z
                \end{array},
              \right. \hspace{1cm}
              P = \left\{
                \begin{array}{ll}
                x \rightarrow -x\\
                y \rightarrow -y\\
                z \rightarrow -z
                \end{array}
              \right..\nonumber
\end{eqnarray} 

\subsection{Generators acting on orbitals}

Considering the orbitals in the $t_{2g}$ manifold in the basis $\Phi^\dagger  = (c_{yz}^\dagger,  c_{xz}^\dagger,c_{xy}^\dagger)$,  the basic operations above can be written as:
\begin{eqnarray}
C_4 &=& \begin{pmatrix}
0 & 1 & 0 \\
-1 & 0 & 0 \\
0 & 0 & -1
\end{pmatrix}, \\ \nonumber
C_{2x}' &=& \begin{pmatrix}
1 & 0 & 0 \\
0 & -1 & 0 \\
0 & 0 & -1
\end{pmatrix}, \\ \nonumber
C_{2d}''  &=& \begin{pmatrix}
0 & -1 & 0 \\
-1 & 0 & 0 \\
0 & 0 & 1
\end{pmatrix}, \\ \nonumber
P &=& \begin{pmatrix}
1 & 0 & 0 \\
0 & 1 & 0 \\
0 & 0 & 1
\end{pmatrix}.
\end{eqnarray}

\subsection{Generators acting on spin}

Considering the spin DOF in the basis $\Phi^\dagger  = (c_{\uparrow}^\dagger,  c_{\downarrow}^\dagger)$,  the basic operations above can be written as:
\begin{eqnarray}
C_4 &=& \frac{\sigma_0- i \sigma_3}{\sqrt{2}},  \\ \nonumber
C_{2x}' &=& i \sigma_1,  \\ \nonumber
C_{2d}''  &=& i  \frac{(\sigma_1+\sigma_2)}{\sqrt{2}}, \\ \nonumber
P &=& \sigma_0.
\end{eqnarray}

\newpage
\section{Irreducible representations of the matrix-basis of the order parameters}\label{Sec:OP}

Here we summarize the properties of the 15 matrices which would pair with even $d_{ab}(\bk)$ functions:
\begin{center}
    \begin{tabular}{| c | c | c | c |}
    \hline
    $[a,b]$ &    Irrep  & Orbital & Spin \\ \hline
    $[0,0]$ &  $A_{1}$ &  Intra & Singlet  \\ \hline
    $[4,3]$ &  $A_{1}$ &  Singlet & Triplet  \\ \hline
    $[8,0]$ &  $A_{1}$ &  Intra & Singlet  \\ \hline
    $[5,2]-[6,1]$ &  $A_{1}$ & Singlet & Triplet   \\ \hline
    $[5,1]+[6,2]$ &  $A_{2}$ & Singlet & Triplet   \\ \hline
    $[7,0]$ &  $B_{1}$ &  Intra & Singlet  \\ \hline
    $[5,2]+[6,1]$ &  $B_{1}$ & Singlet & Triplet   \\ \hline
    $[1,0]$ &  $B_{2}$ &  Triplet & Singlet  \\ \hline
    $[5,1]-[6,2]$ &  $B_{2}$ & Singlet & Triplet   \\ \hline
    $[3,0]$ &  $E (i)$ &  Triplet & Singlet  \\ \hline
    $-[2,0]$ &  $E (i)$ &  Triplet & Singlet  \\ \hline    
    $[4,2]$ &  $E (ii)$ &  Singlet & Triplet  \\ \hline
    $-[4,1]$ &  $E (ii)$ &  Singlet & Triplet  \\ \hline
    $[5,3]$ &  $E (iii)$ &  Singlet & Triplet  \\ \hline
    $[6,3]$ &  $E (iii)$ &  Singlet & Triplet  \\ \hline
    \end{tabular}
\end{center}

\newpage
Here we summarize the properties of the 21 matrices which would pair with odd $d_{ab}(\bk)$ functions:
\begin{center}
    \begin{tabular}{| c | c | c | c |}
    \hline
    $[a,b]$ &    Irrep  & Orbital & Spin \\ \hline
    $[2,2]-[3,1]$ &  $A_{1}$ & Triplet & Triplet   \\ \hline
    $[0,3]$ &  $A_{2}$ &  Intra & Triplet  \\ \hline
    $[4,0]$ &  $A_{2}$ &  Singlet & Singlet  \\ \hline
    $[8,3]$ &  $A_{2}$ &  Intra & Triplet  \\ \hline    
    $[7,3]$ &  $B_{2}$ &  Intra & Triplet  \\ \hline        
    $[2,1]+[3,2]$ &  $A_{2}$ & Triplet & Triplet   \\ \hline
    $[1,3]$ &  $B_{1}$ &  Intra & Triplet  \\ \hline     
    $[2,2]+[3,1]$ &  $B_{1}$ & Triplet & Triplet   \\ \hline    
    $[2,1]-[3,2]$ &  $B_{2}$ & Triplet & Triplet   \\ \hline
    $[0,1]$ &  $E (i)$ &  Intra & Triplet  \\ \hline
    $[0,2]$ &  $E (i)$ &  Intra & Triplet  \\ \hline
    $[1,2]$ &  $E (ii)$ &  Triplet & Triplet  \\ \hline
    $[1,1]$ &  $E (ii)$ &  Triplet & Triplet  \\ \hline       
    $[2,3]$ &  $E (iii)$ &  Triplet & Triplet  \\ \hline
    $[3,3]$ &  $E (iii)$ &  Triplet & Triplet  \\ \hline    
    $[6,0]$ &  $E (iv)$ &  Singlet & Singlet \\ \hline
    $-[5,0]$ &  $E (iv)$ & Singlet & Singlet  \\ \hline    
    $[7,1]$ &  $E (v)$ &   Intra & Triplet  \\ \hline
    $-[7,2]$ &  $E (v)$ &  Intra & Triplet   \\ \hline
    $[8,1]$ &  $E (vi)$ &   Intra & Triplet  \\ \hline
    $[8,2]$ &  $E (vi)$ &  Intra & Triplet   \\ \hline      
    \end{tabular}
\end{center}

\newpage. 

\newpage

\bibliographystyle{apsrev4-1}
\bibliography{SROPR}{}

\begin{thebibliography}{51}%
\makeatletter
\providecommand \@ifxundefined [1]{%
 \@ifx{#1\undefined}
}%
\providecommand \@ifnum [1]{%
 \ifnum #1\expandafter \@firstoftwo
 \else \expandafter \@secondoftwo
 \fi
}%
\providecommand \@ifx [1]{%
 \ifx #1\expandafter \@firstoftwo
 \else \expandafter \@secondoftwo
 \fi
}%
\providecommand \natexlab [1]{#1}%
\providecommand \enquote  [1]{``#1''}%
\providecommand \bibnamefont  [1]{#1}%
\providecommand \bibfnamefont [1]{#1}%
\providecommand \citenamefont [1]{#1}%
\providecommand \href@noop [0]{\@secondoftwo}%
\providecommand \href [0]{\begingroup \@sanitize@url \@href}%
\providecommand \@href[1]{\@@startlink{#1}\@@href}%
\providecommand \@@href[1]{\endgroup#1\@@endlink}%
\providecommand \@sanitize@url [0]{\catcode `\\12\catcode `\$12\catcode
  `\&12\catcode `\#12\catcode `\^12\catcode `\_12\catcode `\%12\relax}%
\providecommand \@@startlink[1]{}%
\providecommand \@@endlink[0]{}%
\providecommand \url  [0]{\begingroup\@sanitize@url \@url }%
\providecommand \@url [1]{\endgroup\@href {#1}{\urlprefix }}%
\providecommand \urlprefix  [0]{URL }%
\providecommand \Eprint [0]{\href }%
\providecommand \doibase [0]{http://dx.doi.org/}%
\providecommand \selectlanguage [0]{\@gobble}%
\providecommand \bibinfo  [0]{\@secondoftwo}%
\providecommand \bibfield  [0]{\@secondoftwo}%
\providecommand \translation [1]{[#1]}%
\providecommand \BibitemOpen [0]{}%
\providecommand \bibitemStop [0]{}%
\providecommand \bibitemNoStop [0]{.\EOS\space}%
\providecommand \EOS [0]{\spacefactor3000\relax}%
\providecommand \BibitemShut  [1]{\csname bibitem#1\endcsname}%
\let\auto@bib@innerbib\@empty
\bibitem [{\citenamefont {Mackenzie}\ and\ \citenamefont
  {Maeno}(2003)}]{MacRMP03}%
  \BibitemOpen
  \bibfield  {author} {\bibinfo {author} {\bibfnamefont {A.~P.}\ \bibnamefont
  {Mackenzie}}\ and\ \bibinfo {author} {\bibfnamefont {Y.}~\bibnamefont
  {Maeno}},\ }\href {\doibase 10.1103/RevModPhys.75.657} {\bibfield  {journal}
  {\bibinfo  {journal} {Rev. Mod. Phys.}\ }\textbf {\bibinfo {volume} {75}},\
  \bibinfo {pages} {657} (\bibinfo {year} {2003})}\BibitemShut {NoStop}%
\bibitem [{\citenamefont {Mao}\ \emph {et~al.}(2000)\citenamefont {Mao},
  \citenamefont {Maeno},\ and\ \citenamefont {Fukazawa}}]{MaoMRB00}%
  \BibitemOpen
  \bibfield  {author} {\bibinfo {author} {\bibfnamefont {Z.}~\bibnamefont
  {Mao}}, \bibinfo {author} {\bibfnamefont {Y.}~\bibnamefont {Maeno}}, \ and\
  \bibinfo {author} {\bibfnamefont {H.}~\bibnamefont {Fukazawa}},\ }\href
  {\doibase https://doi.org/10.1016/S0025-5408(00)00378-0} {\bibfield
  {journal} {\bibinfo  {journal} {Materials Research Bulletin}\ }\textbf
  {\bibinfo {volume} {35}},\ \bibinfo {pages} {1813 } (\bibinfo {year}
  {2000})}\BibitemShut {NoStop}%
\bibitem [{\citenamefont {Mackenzie}\ \emph {et~al.}(1996)\citenamefont
  {Mackenzie}, \citenamefont {Julian}, \citenamefont {Diver}, \citenamefont
  {McMullan}, \citenamefont {Ray}, \citenamefont {Lonzarich}, \citenamefont
  {Maeno}, \citenamefont {Nishizaki},\ and\ \citenamefont {Fujita}}]{MacPRL96}%
  \BibitemOpen
  \bibfield  {author} {\bibinfo {author} {\bibfnamefont {A.~P.}\ \bibnamefont
  {Mackenzie}}, \bibinfo {author} {\bibfnamefont {S.~R.}\ \bibnamefont
  {Julian}}, \bibinfo {author} {\bibfnamefont {A.~J.}\ \bibnamefont {Diver}},
  \bibinfo {author} {\bibfnamefont {G.~J.}\ \bibnamefont {McMullan}}, \bibinfo
  {author} {\bibfnamefont {M.~P.}\ \bibnamefont {Ray}}, \bibinfo {author}
  {\bibfnamefont {G.~G.}\ \bibnamefont {Lonzarich}}, \bibinfo {author}
  {\bibfnamefont {Y.}~\bibnamefont {Maeno}}, \bibinfo {author} {\bibfnamefont
  {S.}~\bibnamefont {Nishizaki}}, \ and\ \bibinfo {author} {\bibfnamefont
  {T.}~\bibnamefont {Fujita}},\ }\href {\doibase 10.1103/PhysRevLett.76.3786}
  {\bibfield  {journal} {\bibinfo  {journal} {Phys. Rev. Lett.}\ }\textbf
  {\bibinfo {volume} {76}},\ \bibinfo {pages} {3786} (\bibinfo {year}
  {1996})}\BibitemShut {NoStop}%
\bibitem [{\citenamefont {Bergemann}\ \emph {et~al.}(2000)\citenamefont
  {Bergemann}, \citenamefont {Julian}, \citenamefont {Mackenzie}, \citenamefont
  {Nishizaki},\ and\ \citenamefont {Maeno}}]{BerPRL00}%
  \BibitemOpen
  \bibfield  {author} {\bibinfo {author} {\bibfnamefont {C.}~\bibnamefont
  {Bergemann}}, \bibinfo {author} {\bibfnamefont {S.~R.}\ \bibnamefont
  {Julian}}, \bibinfo {author} {\bibfnamefont {A.~P.}\ \bibnamefont
  {Mackenzie}}, \bibinfo {author} {\bibfnamefont {S.}~\bibnamefont
  {Nishizaki}}, \ and\ \bibinfo {author} {\bibfnamefont {Y.}~\bibnamefont
  {Maeno}},\ }\href {\doibase 10.1103/PhysRevLett.84.2662} {\bibfield
  {journal} {\bibinfo  {journal} {Phys. Rev. Lett.}\ }\textbf {\bibinfo
  {volume} {84}},\ \bibinfo {pages} {2662} (\bibinfo {year}
  {2000})}\BibitemShut {NoStop}%
\bibitem [{\citenamefont {Maeno}\ \emph {et~al.}(1997)\citenamefont {Maeno},
  \citenamefont {Yoshida}, \citenamefont {Hashimoto}, \citenamefont
  {Nishizaki}, \citenamefont {Ikeda}, \citenamefont {Nohara}, \citenamefont
  {Fujita}, \citenamefont {Mackenzie}, \citenamefont {Hussey}, \citenamefont
  {Bednorz},\ and\ \citenamefont {Lichtenberg}}]{MaeJPSJ97}%
  \BibitemOpen
  \bibfield  {author} {\bibinfo {author} {\bibfnamefont {Y.}~\bibnamefont
  {Maeno}}, \bibinfo {author} {\bibfnamefont {K.}~\bibnamefont {Yoshida}},
  \bibinfo {author} {\bibfnamefont {H.}~\bibnamefont {Hashimoto}}, \bibinfo
  {author} {\bibfnamefont {S.}~\bibnamefont {Nishizaki}}, \bibinfo {author}
  {\bibfnamefont {S.-i.}\ \bibnamefont {Ikeda}}, \bibinfo {author}
  {\bibfnamefont {M.}~\bibnamefont {Nohara}}, \bibinfo {author} {\bibfnamefont
  {T.}~\bibnamefont {Fujita}}, \bibinfo {author} {\bibfnamefont
  {A.}~\bibnamefont {Mackenzie}}, \bibinfo {author} {\bibfnamefont
  {N.}~\bibnamefont {Hussey}}, \bibinfo {author} {\bibfnamefont
  {J.}~\bibnamefont {Bednorz}}, \ and\ \bibinfo {author} {\bibfnamefont
  {F.}~\bibnamefont {Lichtenberg}},\ }\href {\doibase 10.1143/JPSJ.66.1405}
  {\bibfield  {journal} {\bibinfo  {journal} {Journal of the Physical Society
  of Japan}\ }\textbf {\bibinfo {volume} {66}},\ \bibinfo {pages} {1405}
  (\bibinfo {year} {1997})},\ \Eprint
  {http://arxiv.org/abs/https://doi.org/10.1143/JPSJ.66.1405}
  {https://doi.org/10.1143/JPSJ.66.1405} \BibitemShut {NoStop}%
\bibitem [{\citenamefont {Damascelli}\ \emph {et~al.}(2000)\citenamefont
  {Damascelli}, \citenamefont {Lu}, \citenamefont {Shen}, \citenamefont
  {Armitage}, \citenamefont {Ronning}, \citenamefont {Feng}, \citenamefont
  {Kim}, \citenamefont {Shen}, \citenamefont {Kimura}, \citenamefont {Tokura},
  \citenamefont {Mao},\ and\ \citenamefont {Maeno}}]{DamPRL00}%
  \BibitemOpen
  \bibfield  {author} {\bibinfo {author} {\bibfnamefont {A.}~\bibnamefont
  {Damascelli}}, \bibinfo {author} {\bibfnamefont {D.~H.}\ \bibnamefont {Lu}},
  \bibinfo {author} {\bibfnamefont {K.~M.}\ \bibnamefont {Shen}}, \bibinfo
  {author} {\bibfnamefont {N.~P.}\ \bibnamefont {Armitage}}, \bibinfo {author}
  {\bibfnamefont {F.}~\bibnamefont {Ronning}}, \bibinfo {author} {\bibfnamefont
  {D.~L.}\ \bibnamefont {Feng}}, \bibinfo {author} {\bibfnamefont
  {C.}~\bibnamefont {Kim}}, \bibinfo {author} {\bibfnamefont {Z.-X.}\
  \bibnamefont {Shen}}, \bibinfo {author} {\bibfnamefont {T.}~\bibnamefont
  {Kimura}}, \bibinfo {author} {\bibfnamefont {Y.}~\bibnamefont {Tokura}},
  \bibinfo {author} {\bibfnamefont {Z.~Q.}\ \bibnamefont {Mao}}, \ and\
  \bibinfo {author} {\bibfnamefont {Y.}~\bibnamefont {Maeno}},\ }\href
  {\doibase 10.1103/PhysRevLett.85.5194} {\bibfield  {journal} {\bibinfo
  {journal} {Phys. Rev. Lett.}\ }\textbf {\bibinfo {volume} {85}},\ \bibinfo
  {pages} {5194} (\bibinfo {year} {2000})}\BibitemShut {NoStop}%
\bibitem [{\citenamefont {Mackenzie}\ \emph {et~al.}(2017)\citenamefont
  {Mackenzie}, \citenamefont {Scaffidi}, \citenamefont {Hicks},\ and\
  \citenamefont {Maeno}}]{MacNJP17}%
  \BibitemOpen
  \bibfield  {author} {\bibinfo {author} {\bibfnamefont {A.~P.}\ \bibnamefont
  {Mackenzie}}, \bibinfo {author} {\bibfnamefont {T.}~\bibnamefont {Scaffidi}},
  \bibinfo {author} {\bibfnamefont {C.~W.}\ \bibnamefont {Hicks}}, \ and\
  \bibinfo {author} {\bibfnamefont {Y.}~\bibnamefont {Maeno}},\ }\href
  {\doibase 10.1038/s41535-017-0045-4} {\bibfield  {journal} {\bibinfo
  {journal} {npj Quantum Materials}\ }\textbf {\bibinfo {volume} {2}},\
  \bibinfo {pages} {40} (\bibinfo {year} {2017})}\BibitemShut {NoStop}%
\bibitem [{\citenamefont {Nomura}\ and\ \citenamefont
  {Yamada}(2000)}]{NomJPSJ00}%
  \BibitemOpen
  \bibfield  {author} {\bibinfo {author} {\bibfnamefont {T.}~\bibnamefont
  {Nomura}}\ and\ \bibinfo {author} {\bibfnamefont {K.}~\bibnamefont
  {Yamada}},\ }\href {\doibase 10.1143/JPSJ.69.3678} {\bibfield  {journal}
  {\bibinfo  {journal} {Journal of the Physical Society of Japan}\ }\textbf
  {\bibinfo {volume} {69}},\ \bibinfo {pages} {3678} (\bibinfo {year}
  {2000})},\ \Eprint
  {http://arxiv.org/abs/https://doi.org/10.1143/JPSJ.69.3678}
  {https://doi.org/10.1143/JPSJ.69.3678} \BibitemShut {NoStop}%
\bibitem [{\citenamefont {Nomura}\ and\ \citenamefont
  {Yamada}(2002)}]{NomJPSJ02}%
  \BibitemOpen
  \bibfield  {author} {\bibinfo {author} {\bibfnamefont {T.}~\bibnamefont
  {Nomura}}\ and\ \bibinfo {author} {\bibfnamefont {K.}~\bibnamefont
  {Yamada}},\ }\href {\doibase 10.1143/JPSJ.71.1993} {\bibfield  {journal}
  {\bibinfo  {journal} {Journal of the Physical Society of Japan}\ }\textbf
  {\bibinfo {volume} {71}},\ \bibinfo {pages} {1993} (\bibinfo {year}
  {2002})},\ \Eprint
  {http://arxiv.org/abs/https://doi.org/10.1143/JPSJ.71.1993}
  {https://doi.org/10.1143/JPSJ.71.1993} \BibitemShut {NoStop}%
\bibitem [{\citenamefont {Miyake}\ and\ \citenamefont
  {Narikiyo}(1999)}]{MiyPRL99}%
  \BibitemOpen
  \bibfield  {author} {\bibinfo {author} {\bibfnamefont {K.}~\bibnamefont
  {Miyake}}\ and\ \bibinfo {author} {\bibfnamefont {O.}~\bibnamefont
  {Narikiyo}},\ }\href {\doibase 10.1103/PhysRevLett.83.1423} {\bibfield
  {journal} {\bibinfo  {journal} {Phys. Rev. Lett.}\ }\textbf {\bibinfo
  {volume} {83}},\ \bibinfo {pages} {1423} (\bibinfo {year}
  {1999})}\BibitemShut {NoStop}%
\bibitem [{\citenamefont {Kuwabara}\ and\ \citenamefont
  {Ogata}(2000)}]{KuwPRL00}%
  \BibitemOpen
  \bibfield  {author} {\bibinfo {author} {\bibfnamefont {T.}~\bibnamefont
  {Kuwabara}}\ and\ \bibinfo {author} {\bibfnamefont {M.}~\bibnamefont
  {Ogata}},\ }\href {\doibase 10.1103/PhysRevLett.85.4586} {\bibfield
  {journal} {\bibinfo  {journal} {Phys. Rev. Lett.}\ }\textbf {\bibinfo
  {volume} {85}},\ \bibinfo {pages} {4586} (\bibinfo {year}
  {2000})}\BibitemShut {NoStop}%
\bibitem [{\citenamefont {Ishida}\ \emph {et~al.}(1998)\citenamefont {Ishida},
  \citenamefont {Mukuda}, \citenamefont {Kitaoka}, \citenamefont {Asayama},
  \citenamefont {Mao}, \citenamefont {Mori},\ and\ \citenamefont
  {Maeno}}]{IshNat98}%
  \BibitemOpen
  \bibfield  {author} {\bibinfo {author} {\bibfnamefont {K.}~\bibnamefont
  {Ishida}}, \bibinfo {author} {\bibfnamefont {H.}~\bibnamefont {Mukuda}},
  \bibinfo {author} {\bibfnamefont {Y.}~\bibnamefont {Kitaoka}}, \bibinfo
  {author} {\bibfnamefont {K.}~\bibnamefont {Asayama}}, \bibinfo {author}
  {\bibfnamefont {Z.~Q.}\ \bibnamefont {Mao}}, \bibinfo {author} {\bibfnamefont
  {Y.}~\bibnamefont {Mori}}, \ and\ \bibinfo {author} {\bibfnamefont
  {Y.}~\bibnamefont {Maeno}},\ }\href {http://dx.doi.org/10.1038/25315}
  {\bibfield  {journal} {\bibinfo  {journal} {Nature}\ }\textbf {\bibinfo
  {volume} {396}},\ \bibinfo {pages} {658 EP } (\bibinfo {year}
  {1998})}\BibitemShut {NoStop}%
\bibitem [{\citenamefont {Ishida}\ \emph {et~al.}(2015)\citenamefont {Ishida},
  \citenamefont {Manago}, \citenamefont {Yamanaka}, \citenamefont {Fukazawa},
  \citenamefont {Mao}, \citenamefont {Maeno},\ and\ \citenamefont
  {Miyake}}]{IshPRB15}%
  \BibitemOpen
  \bibfield  {author} {\bibinfo {author} {\bibfnamefont {K.}~\bibnamefont
  {Ishida}}, \bibinfo {author} {\bibfnamefont {M.}~\bibnamefont {Manago}},
  \bibinfo {author} {\bibfnamefont {T.}~\bibnamefont {Yamanaka}}, \bibinfo
  {author} {\bibfnamefont {H.}~\bibnamefont {Fukazawa}}, \bibinfo {author}
  {\bibfnamefont {Z.~Q.}\ \bibnamefont {Mao}}, \bibinfo {author} {\bibfnamefont
  {Y.}~\bibnamefont {Maeno}}, \ and\ \bibinfo {author} {\bibfnamefont
  {K.}~\bibnamefont {Miyake}},\ }\href {\doibase 10.1103/PhysRevB.92.100502}
  {\bibfield  {journal} {\bibinfo  {journal} {Phys. Rev. B}\ }\textbf {\bibinfo
  {volume} {92}},\ \bibinfo {pages} {100502(R)} (\bibinfo {year}
  {2015})}\BibitemShut {NoStop}%
\bibitem [{\citenamefont {Duffy}\ \emph {et~al.}(2000)\citenamefont {Duffy},
  \citenamefont {Hayden}, \citenamefont {Maeno}, \citenamefont {Mao},
  \citenamefont {Kulda},\ and\ \citenamefont {McIntyre}}]{DufPRL00}%
  \BibitemOpen
  \bibfield  {author} {\bibinfo {author} {\bibfnamefont {J.~A.}\ \bibnamefont
  {Duffy}}, \bibinfo {author} {\bibfnamefont {S.~M.}\ \bibnamefont {Hayden}},
  \bibinfo {author} {\bibfnamefont {Y.}~\bibnamefont {Maeno}}, \bibinfo
  {author} {\bibfnamefont {Z.}~\bibnamefont {Mao}}, \bibinfo {author}
  {\bibfnamefont {J.}~\bibnamefont {Kulda}}, \ and\ \bibinfo {author}
  {\bibfnamefont {G.~J.}\ \bibnamefont {McIntyre}},\ }\href {\doibase
  10.1103/PhysRevLett.85.5412} {\bibfield  {journal} {\bibinfo  {journal}
  {Phys. Rev. Lett.}\ }\textbf {\bibinfo {volume} {85}},\ \bibinfo {pages}
  {5412} (\bibinfo {year} {2000})}\BibitemShut {NoStop}%
\bibitem [{\citenamefont {Luke}\ \emph {et~al.}(1998)\citenamefont {Luke},
  \citenamefont {Fudamoto}, \citenamefont {Kojima}, \citenamefont {Larkin},
  \citenamefont {Merrin}, \citenamefont {Nachumi}, \citenamefont {Uemura},
  \citenamefont {Maeno}, \citenamefont {Mao}, \citenamefont {Mori},
  \citenamefont {Nakamura},\ and\ \citenamefont {Sigrist}}]{LukNat98}%
  \BibitemOpen
  \bibfield  {author} {\bibinfo {author} {\bibfnamefont {G.~M.}\ \bibnamefont
  {Luke}}, \bibinfo {author} {\bibfnamefont {Y.}~\bibnamefont {Fudamoto}},
  \bibinfo {author} {\bibfnamefont {K.~M.}\ \bibnamefont {Kojima}}, \bibinfo
  {author} {\bibfnamefont {M.~I.}\ \bibnamefont {Larkin}}, \bibinfo {author}
  {\bibfnamefont {J.}~\bibnamefont {Merrin}}, \bibinfo {author} {\bibfnamefont
  {B.}~\bibnamefont {Nachumi}}, \bibinfo {author} {\bibfnamefont {Y.~J.}\
  \bibnamefont {Uemura}}, \bibinfo {author} {\bibfnamefont {Y.}~\bibnamefont
  {Maeno}}, \bibinfo {author} {\bibfnamefont {Z.~Q.}\ \bibnamefont {Mao}},
  \bibinfo {author} {\bibfnamefont {Y.}~\bibnamefont {Mori}}, \bibinfo {author}
  {\bibfnamefont {H.}~\bibnamefont {Nakamura}}, \ and\ \bibinfo {author}
  {\bibfnamefont {M.}~\bibnamefont {Sigrist}},\ }\href
  {http://dx.doi.org/10.1038/29038} {\bibfield  {journal} {\bibinfo  {journal}
  {Nature}\ }\textbf {\bibinfo {volume} {394}},\ \bibinfo {pages} {558 EP }
  (\bibinfo {year} {1998})}\BibitemShut {NoStop}%
\bibitem [{\citenamefont {Luke}\ \emph {et~al.}(2000)\citenamefont {Luke},
  \citenamefont {Fudamoto}, \citenamefont {Kojima}, \citenamefont {Larkin},
  \citenamefont {Nachumi}, \citenamefont {Uemura}, \citenamefont {Sonier},
  \citenamefont {Maeno}, \citenamefont {Mao}, \citenamefont {Mori},\ and\
  \citenamefont {Agterberg}}]{LukPB00}%
  \BibitemOpen
  \bibfield  {author} {\bibinfo {author} {\bibfnamefont {G.}~\bibnamefont
  {Luke}}, \bibinfo {author} {\bibfnamefont {Y.}~\bibnamefont {Fudamoto}},
  \bibinfo {author} {\bibfnamefont {K.}~\bibnamefont {Kojima}}, \bibinfo
  {author} {\bibfnamefont {M.}~\bibnamefont {Larkin}}, \bibinfo {author}
  {\bibfnamefont {B.}~\bibnamefont {Nachumi}}, \bibinfo {author} {\bibfnamefont
  {Y.}~\bibnamefont {Uemura}}, \bibinfo {author} {\bibfnamefont
  {J.}~\bibnamefont {Sonier}}, \bibinfo {author} {\bibfnamefont
  {Y.}~\bibnamefont {Maeno}}, \bibinfo {author} {\bibfnamefont
  {Z.}~\bibnamefont {Mao}}, \bibinfo {author} {\bibfnamefont {Y.}~\bibnamefont
  {Mori}}, \ and\ \bibinfo {author} {\bibfnamefont {D.}~\bibnamefont
  {Agterberg}},\ }\href {\doibase
  https://doi.org/10.1016/S0921-4526(00)00414-2} {\bibfield  {journal}
  {\bibinfo  {journal} {Physica B: Condensed Matter}\ }\textbf {\bibinfo
  {volume} {289-290}},\ \bibinfo {pages} {373 } (\bibinfo {year}
  {2000})}\BibitemShut {NoStop}%
\bibitem [{\citenamefont {Xia}\ \emph {et~al.}(2006)\citenamefont {Xia},
  \citenamefont {Maeno}, \citenamefont {Beyersdorf}, \citenamefont {Fejer},\
  and\ \citenamefont {Kapitulnik}}]{XiaPRL06}%
  \BibitemOpen
  \bibfield  {author} {\bibinfo {author} {\bibfnamefont {J.}~\bibnamefont
  {Xia}}, \bibinfo {author} {\bibfnamefont {Y.}~\bibnamefont {Maeno}}, \bibinfo
  {author} {\bibfnamefont {P.~T.}\ \bibnamefont {Beyersdorf}}, \bibinfo
  {author} {\bibfnamefont {M.~M.}\ \bibnamefont {Fejer}}, \ and\ \bibinfo
  {author} {\bibfnamefont {A.}~\bibnamefont {Kapitulnik}},\ }\href {\doibase
  10.1103/PhysRevLett.97.167002} {\bibfield  {journal} {\bibinfo  {journal}
  {Phys. Rev. Lett.}\ }\textbf {\bibinfo {volume} {97}},\ \bibinfo {pages}
  {167002} (\bibinfo {year} {2006})}\BibitemShut {NoStop}%
\bibitem [{\citenamefont {Rice}\ and\ \citenamefont
  {Sigrist}(1995)}]{RicJPCM1995}%
  \BibitemOpen
  \bibfield  {author} {\bibinfo {author} {\bibfnamefont {T.~M.}\ \bibnamefont
  {Rice}}\ and\ \bibinfo {author} {\bibfnamefont {M.}~\bibnamefont {Sigrist}},\
  }\href {http://stacks.iop.org/0953-8984/7/i=47/a=002} {\bibfield  {journal}
  {\bibinfo  {journal} {Journal of Physics: Condensed Matter}\ }\textbf
  {\bibinfo {volume} {7}},\ \bibinfo {pages} {L643} (\bibinfo {year}
  {1995})}\BibitemShut {NoStop}%
\bibitem [{\citenamefont {Maeno}\ \emph {et~al.}(2012)\citenamefont {Maeno},
  \citenamefont {Kittaka}, \citenamefont {Nomura}, \citenamefont {Yonezawa},\
  and\ \citenamefont {Ishida}}]{MaeJPSJ12}%
  \BibitemOpen
  \bibfield  {author} {\bibinfo {author} {\bibfnamefont {Y.}~\bibnamefont
  {Maeno}}, \bibinfo {author} {\bibfnamefont {S.}~\bibnamefont {Kittaka}},
  \bibinfo {author} {\bibfnamefont {T.}~\bibnamefont {Nomura}}, \bibinfo
  {author} {\bibfnamefont {S.}~\bibnamefont {Yonezawa}}, \ and\ \bibinfo
  {author} {\bibfnamefont {K.}~\bibnamefont {Ishida}},\ }\href {\doibase
  10.1143/JPSJ.81.011009} {\bibfield  {journal} {\bibinfo  {journal} {Journal
  of the Physical Society of Japan}\ }\textbf {\bibinfo {volume} {81}},\
  \bibinfo {pages} {011009} (\bibinfo {year} {2012})},\ \Eprint
  {http://arxiv.org/abs/https://doi.org/10.1143/JPSJ.81.011009}
  {https://doi.org/10.1143/JPSJ.81.011009} \BibitemShut {NoStop}%
\bibitem [{\citenamefont {Kallin}(2012)}]{KalRPP12}%
  \BibitemOpen
  \bibfield  {author} {\bibinfo {author} {\bibfnamefont {C.}~\bibnamefont
  {Kallin}},\ }\href {\doibase 10.1088/0034-4885/75/4/042501} {\bibfield
  {journal} {\bibinfo  {journal} {Reports on Progress in Physics}\ }\textbf
  {\bibinfo {volume} {75}} (\bibinfo {year} {2012}),\
  10.1088/0034-4885/75/4/042501}\BibitemShut {NoStop}%
\bibitem [{\citenamefont {NishiZaki}\ \emph {et~al.}(1999)\citenamefont
  {NishiZaki}, \citenamefont {Maeno},\ and\ \citenamefont {Mao}}]{NisJLTP99}%
  \BibitemOpen
  \bibfield  {author} {\bibinfo {author} {\bibfnamefont {S.}~\bibnamefont
  {NishiZaki}}, \bibinfo {author} {\bibfnamefont {Y.}~\bibnamefont {Maeno}}, \
  and\ \bibinfo {author} {\bibfnamefont {Z.}~\bibnamefont {Mao}},\ }\href
  {\doibase 10.1023/A:1022551313401} {\bibfield  {journal} {\bibinfo  {journal}
  {Journal of Low Temperature Physics}\ }\textbf {\bibinfo {volume} {117}},\
  \bibinfo {pages} {1581} (\bibinfo {year} {1999})}\BibitemShut {NoStop}%
\bibitem [{\citenamefont {NishiZaki}\ \emph {et~al.}(2000)\citenamefont
  {NishiZaki}, \citenamefont {Maeno},\ and\ \citenamefont {Mao}}]{NisJPSJ00}%
  \BibitemOpen
  \bibfield  {author} {\bibinfo {author} {\bibfnamefont {S.}~\bibnamefont
  {NishiZaki}}, \bibinfo {author} {\bibfnamefont {Y.}~\bibnamefont {Maeno}}, \
  and\ \bibinfo {author} {\bibfnamefont {Z.}~\bibnamefont {Mao}},\ }\href
  {\doibase 10.1143/JPSJ.69.572} {\bibfield  {journal} {\bibinfo  {journal}
  {Journal of the Physical Society of Japan}\ }\textbf {\bibinfo {volume}
  {69}},\ \bibinfo {pages} {572} (\bibinfo {year} {2000})},\ \Eprint
  {http://arxiv.org/abs/https://doi.org/10.1143/JPSJ.69.572}
  {https://doi.org/10.1143/JPSJ.69.572} \BibitemShut {NoStop}%
\bibitem [{\citenamefont {Deguchi}\ \emph
  {et~al.}(2004{\natexlab{a}})\citenamefont {Deguchi}, \citenamefont {Mao},
  \citenamefont {Yaguchi},\ and\ \citenamefont {Maeno}}]{DegPRL04}%
  \BibitemOpen
  \bibfield  {author} {\bibinfo {author} {\bibfnamefont {K.}~\bibnamefont
  {Deguchi}}, \bibinfo {author} {\bibfnamefont {Z.~Q.}\ \bibnamefont {Mao}},
  \bibinfo {author} {\bibfnamefont {H.}~\bibnamefont {Yaguchi}}, \ and\
  \bibinfo {author} {\bibfnamefont {Y.}~\bibnamefont {Maeno}},\ }\href
  {\doibase 10.1103/PhysRevLett.92.047002} {\bibfield  {journal} {\bibinfo
  {journal} {Phys. Rev. Lett.}\ }\textbf {\bibinfo {volume} {92}},\ \bibinfo
  {pages} {047002} (\bibinfo {year} {2004}{\natexlab{a}})}\BibitemShut
  {NoStop}%
\bibitem [{\citenamefont {Deguchi}\ \emph
  {et~al.}(2004{\natexlab{b}})\citenamefont {Deguchi}, \citenamefont {Q.~Mao},\
  and\ \citenamefont {Maeno}}]{DegJPSJ04}%
  \BibitemOpen
  \bibfield  {author} {\bibinfo {author} {\bibfnamefont {K.}~\bibnamefont
  {Deguchi}}, \bibinfo {author} {\bibfnamefont {Z.}~\bibnamefont {Q.~Mao}}, \
  and\ \bibinfo {author} {\bibfnamefont {Y.}~\bibnamefont {Maeno}},\ }\href
  {\doibase 10.1143/JPSJ.73.1313} {\bibfield  {journal} {\bibinfo  {journal}
  {Journal of the Physical Society of Japan}\ }\textbf {\bibinfo {volume}
  {73}},\ \bibinfo {pages} {1313} (\bibinfo {year} {2004}{\natexlab{b}})},\
  \Eprint {http://arxiv.org/abs/https://doi.org/10.1143/JPSJ.73.1313}
  {https://doi.org/10.1143/JPSJ.73.1313} \BibitemShut {NoStop}%
\bibitem [{\citenamefont {Kittaka}\ \emph {et~al.}(2018)\citenamefont
  {Kittaka}, \citenamefont {Nakamura}, \citenamefont {Sakakibara},
  \citenamefont {Kikugawa}, \citenamefont {Terashima}, \citenamefont {Uji},
  \citenamefont {Sokolov}, \citenamefont {Mackenzie}, \citenamefont {Irie},
  \citenamefont {Tsutsumi}, \citenamefont {Suzuki},\ and\ \citenamefont
  {Machida}}]{Kit18}%
  \BibitemOpen
  \bibfield  {author} {\bibinfo {author} {\bibfnamefont {S.}~\bibnamefont
  {Kittaka}}, \bibinfo {author} {\bibfnamefont {S.}~\bibnamefont {Nakamura}},
  \bibinfo {author} {\bibfnamefont {T.}~\bibnamefont {Sakakibara}}, \bibinfo
  {author} {\bibfnamefont {N.}~\bibnamefont {Kikugawa}}, \bibinfo {author}
  {\bibfnamefont {T.}~\bibnamefont {Terashima}}, \bibinfo {author}
  {\bibfnamefont {S.}~\bibnamefont {Uji}}, \bibinfo {author} {\bibfnamefont
  {D.~A.}\ \bibnamefont {Sokolov}}, \bibinfo {author} {\bibfnamefont {A.~P.}\
  \bibnamefont {Mackenzie}}, \bibinfo {author} {\bibfnamefont {K.}~\bibnamefont
  {Irie}}, \bibinfo {author} {\bibfnamefont {Y.}~\bibnamefont {Tsutsumi}},
  \bibinfo {author} {\bibfnamefont {K.}~\bibnamefont {Suzuki}}, \ and\ \bibinfo
  {author} {\bibfnamefont {K.}~\bibnamefont {Machida}},\ }\href {\doibase
  10.7566/JPSJ.87.093703} {\bibfield  {journal} {\bibinfo  {journal} {Journal
  of the Physical Society of Japan}\ }\textbf {\bibinfo {volume} {87}},\
  \bibinfo {pages} {093703} (\bibinfo {year} {2018})}\BibitemShut {NoStop}%
\bibitem [{\citenamefont {Lupien}\ \emph {et~al.}(2001)\citenamefont {Lupien},
  \citenamefont {MacFarlane}, \citenamefont {Proust}, \citenamefont
  {Taillefer}, \citenamefont {Mao},\ and\ \citenamefont {Maeno}}]{LupPRL01}%
  \BibitemOpen
  \bibfield  {author} {\bibinfo {author} {\bibfnamefont {C.}~\bibnamefont
  {Lupien}}, \bibinfo {author} {\bibfnamefont {W.~A.}\ \bibnamefont
  {MacFarlane}}, \bibinfo {author} {\bibfnamefont {C.}~\bibnamefont {Proust}},
  \bibinfo {author} {\bibfnamefont {L.}~\bibnamefont {Taillefer}}, \bibinfo
  {author} {\bibfnamefont {Z.~Q.}\ \bibnamefont {Mao}}, \ and\ \bibinfo
  {author} {\bibfnamefont {Y.}~\bibnamefont {Maeno}},\ }\href {\doibase
  10.1103/PhysRevLett.86.5986} {\bibfield  {journal} {\bibinfo  {journal}
  {Phys. Rev. Lett.}\ }\textbf {\bibinfo {volume} {86}},\ \bibinfo {pages}
  {5986} (\bibinfo {year} {2001})}\BibitemShut {NoStop}%
\bibitem [{\citenamefont {Hassinger}\ \emph {et~al.}(2017)\citenamefont
  {Hassinger}, \citenamefont {Bourgeois-Hope}, \citenamefont {Taniguchi},
  \citenamefont {Ren\'e~de Cotret}, \citenamefont {Grissonnanche},
  \citenamefont {Anwar}, \citenamefont {Maeno}, \citenamefont
  {Doiron-Leyraud},\ and\ \citenamefont {Taillefer}}]{HasPRB2017}%
  \BibitemOpen
  \bibfield  {author} {\bibinfo {author} {\bibfnamefont {E.}~\bibnamefont
  {Hassinger}}, \bibinfo {author} {\bibfnamefont {P.}~\bibnamefont
  {Bourgeois-Hope}}, \bibinfo {author} {\bibfnamefont {H.}~\bibnamefont
  {Taniguchi}}, \bibinfo {author} {\bibfnamefont {S.}~\bibnamefont {Ren\'e~de
  Cotret}}, \bibinfo {author} {\bibfnamefont {G.}~\bibnamefont
  {Grissonnanche}}, \bibinfo {author} {\bibfnamefont {M.~S.}\ \bibnamefont
  {Anwar}}, \bibinfo {author} {\bibfnamefont {Y.}~\bibnamefont {Maeno}},
  \bibinfo {author} {\bibfnamefont {N.}~\bibnamefont {Doiron-Leyraud}}, \ and\
  \bibinfo {author} {\bibfnamefont {L.}~\bibnamefont {Taillefer}},\ }\href
  {\doibase 10.1103/PhysRevX.7.011032} {\bibfield  {journal} {\bibinfo
  {journal} {Phys. Rev. X}\ }\textbf {\bibinfo {volume} {7}},\ \bibinfo {pages}
  {011032} (\bibinfo {year} {2017})}\BibitemShut {NoStop}%
\bibitem [{\citenamefont {{Pustogow}}\ \emph {et~al.}(2019)\citenamefont
  {{Pustogow}}, \citenamefont {{Luo}}, \citenamefont {{Chronister}},
  \citenamefont {{Su}}, \citenamefont {{Sokolov}}, \citenamefont
  {{Jerzembeck}}, \citenamefont {{Mackenzie}}, \citenamefont {{Hicks}},
  \citenamefont {{Kikugawa}}, \citenamefont {{Raghu}}, \citenamefont
  {{Bauer}},\ and\ \citenamefont {{Brown}}}]{Pus19}%
  \BibitemOpen
  \bibfield  {author} {\bibinfo {author} {\bibfnamefont {A.}~\bibnamefont
  {{Pustogow}}}, \bibinfo {author} {\bibfnamefont {Y.}~\bibnamefont {{Luo}}},
  \bibinfo {author} {\bibfnamefont {A.}~\bibnamefont {{Chronister}}}, \bibinfo
  {author} {\bibfnamefont {Y.~S.}\ \bibnamefont {{Su}}}, \bibinfo {author}
  {\bibfnamefont {D.~A.}\ \bibnamefont {{Sokolov}}}, \bibinfo {author}
  {\bibfnamefont {F.}~\bibnamefont {{Jerzembeck}}}, \bibinfo {author}
  {\bibfnamefont {A.~P.}\ \bibnamefont {{Mackenzie}}}, \bibinfo {author}
  {\bibfnamefont {C.~W.}\ \bibnamefont {{Hicks}}}, \bibinfo {author}
  {\bibfnamefont {N.}~\bibnamefont {{Kikugawa}}}, \bibinfo {author}
  {\bibfnamefont {S.}~\bibnamefont {{Raghu}}}, \bibinfo {author} {\bibfnamefont
  {E.~D.}\ \bibnamefont {{Bauer}}}, \ and\ \bibinfo {author} {\bibfnamefont
  {S.~E.}\ \bibnamefont {{Brown}}},\ }\href@noop {} {\bibfield  {journal}
  {\bibinfo  {journal} {arXiv e-prints}\ ,\ \bibinfo {eid} {arXiv:1904.00047}}
  (\bibinfo {year} {2019})},\ \Eprint {http://arxiv.org/abs/1904.00047}
  {arXiv:1904.00047 [cond-mat.supr-con]} \BibitemShut {NoStop}%
\bibitem [{\citenamefont {Watson}\ \emph {et~al.}(2018)\citenamefont {Watson},
  \citenamefont {Gibbs}, \citenamefont {Mackenzie}, \citenamefont {Hicks},\
  and\ \citenamefont {Moler}}]{WatPRB18}%
  \BibitemOpen
  \bibfield  {author} {\bibinfo {author} {\bibfnamefont {C.~A.}\ \bibnamefont
  {Watson}}, \bibinfo {author} {\bibfnamefont {A.~S.}\ \bibnamefont {Gibbs}},
  \bibinfo {author} {\bibfnamefont {A.~P.}\ \bibnamefont {Mackenzie}}, \bibinfo
  {author} {\bibfnamefont {C.~W.}\ \bibnamefont {Hicks}}, \ and\ \bibinfo
  {author} {\bibfnamefont {K.~A.}\ \bibnamefont {Moler}},\ }\href {\doibase
  10.1103/PhysRevB.98.094521} {\bibfield  {journal} {\bibinfo  {journal} {Phys.
  Rev. B}\ }\textbf {\bibinfo {volume} {98}},\ \bibinfo {pages} {094521}
  (\bibinfo {year} {2018})}\BibitemShut {NoStop}%
\bibitem [{\citenamefont {Scaffidi}\ \emph {et~al.}(2014)\citenamefont
  {Scaffidi}, \citenamefont {Romers},\ and\ \citenamefont
  {Simon}}]{ScaPRB2014}%
  \BibitemOpen
  \bibfield  {author} {\bibinfo {author} {\bibfnamefont {T.}~\bibnamefont
  {Scaffidi}}, \bibinfo {author} {\bibfnamefont {J.~C.}\ \bibnamefont
  {Romers}}, \ and\ \bibinfo {author} {\bibfnamefont {S.~H.}\ \bibnamefont
  {Simon}},\ }\href {\doibase 10.1103/PhysRevB.89.220510} {\bibfield  {journal}
  {\bibinfo  {journal} {Phys. Rev. B}\ }\textbf {\bibinfo {volume} {89}},\
  \bibinfo {pages} {220510(R)} (\bibinfo {year} {2014})}\BibitemShut {NoStop}%
\bibitem [{\citenamefont {Ramires}\ and\ \citenamefont
  {Sigrist}(2016)}]{RamPRB16}%
  \BibitemOpen
  \bibfield  {author} {\bibinfo {author} {\bibfnamefont {A.}~\bibnamefont
  {Ramires}}\ and\ \bibinfo {author} {\bibfnamefont {M.}~\bibnamefont
  {Sigrist}},\ }\href {\doibase 10.1103/PhysRevB.94.104501} {\bibfield
  {journal} {\bibinfo  {journal} {Phys. Rev. B}\ }\textbf {\bibinfo {volume}
  {94}},\ \bibinfo {pages} {104501} (\bibinfo {year} {2016})}\BibitemShut
  {NoStop}%
\bibitem [{\citenamefont {Ramires}\ \emph {et~al.}(2018)\citenamefont
  {Ramires}, \citenamefont {Agterberg},\ and\ \citenamefont
  {Sigrist}}]{RamPRB18}%
  \BibitemOpen
  \bibfield  {author} {\bibinfo {author} {\bibfnamefont {A.}~\bibnamefont
  {Ramires}}, \bibinfo {author} {\bibfnamefont {D.~F.}\ \bibnamefont
  {Agterberg}}, \ and\ \bibinfo {author} {\bibfnamefont {M.}~\bibnamefont
  {Sigrist}},\ }\href {\doibase 10.1103/PhysRevB.98.024501} {\bibfield
  {journal} {\bibinfo  {journal} {Phys. Rev. B}\ }\textbf {\bibinfo {volume}
  {98}},\ \bibinfo {pages} {024501} (\bibinfo {year} {2018})}\BibitemShut
  {NoStop}%
\bibitem [{\citenamefont {Sigrist}\ and\ \citenamefont
  {Ueda}(1991)}]{Sigrist1991}%
  \BibitemOpen
  \bibfield  {author} {\bibinfo {author} {\bibfnamefont {M.}~\bibnamefont
  {Sigrist}}\ and\ \bibinfo {author} {\bibfnamefont {K.}~\bibnamefont {Ueda}},\
  }\href {\doibase 10.1103/RevModPhys.63.239} {\bibfield  {journal} {\bibinfo
  {journal} {Rev. Mod. Phys.}\ }\textbf {\bibinfo {volume} {63}},\ \bibinfo
  {pages} {239} (\bibinfo {year} {1991})}\BibitemShut {NoStop}%
\bibitem [{\citenamefont {Ramires}\ and\ \citenamefont
  {Sigrist}(2017)}]{Ramires_2017}%
  \BibitemOpen
  \bibfield  {author} {\bibinfo {author} {\bibfnamefont {A.}~\bibnamefont
  {Ramires}}\ and\ \bibinfo {author} {\bibfnamefont {M.}~\bibnamefont
  {Sigrist}},\ }\href {\doibase 10.1088/1742-6596/807/5/052011} {\bibfield
  {journal} {\bibinfo  {journal} {Journal of Physics: Conference Series}\
  }\textbf {\bibinfo {volume} {807}},\ \bibinfo {pages} {052011} (\bibinfo
  {year} {2017})}\BibitemShut {NoStop}%
\bibitem [{\citenamefont {Sigrist}\ and\ \citenamefont
  {E.~Zhitomirsky}(1996)}]{Sigrist96}%
  \BibitemOpen
  \bibfield  {author} {\bibinfo {author} {\bibfnamefont {M.}~\bibnamefont
  {Sigrist}}\ and\ \bibinfo {author} {\bibfnamefont {M.}~\bibnamefont
  {E.~Zhitomirsky}},\ }\href {\doibase 10.1143/JPSJ.65.3452} {\bibfield
  {journal} {\bibinfo  {journal} {Journal of the Physical Society of Japan}\
  }\textbf {\bibinfo {volume} {65}},\ \bibinfo {pages} {3452} (\bibinfo {year}
  {1996})},\ \Eprint
  {http://arxiv.org/abs/https://doi.org/10.1143/JPSJ.65.3452}
  {https://doi.org/10.1143/JPSJ.65.3452} \BibitemShut {NoStop}%
\bibitem [{\citenamefont {Sigrist}(2005)}]{Sigrist05}%
  \BibitemOpen
  \bibfield  {author} {\bibinfo {author} {\bibfnamefont {M.}~\bibnamefont
  {Sigrist}},\ }\href {\doibase 10.1143/PTPS.160.1} {\bibfield  {journal}
  {\bibinfo  {journal} {Progress of Theoretical Physics Supplement}\ }\textbf
  {\bibinfo {volume} {160}},\ \bibinfo {pages} {1} (\bibinfo {year} {2005})},\
  \Eprint
  {http://arxiv.org/abs/http://oup.prod.sis.lan/ptps/article-pdf/doi/10.1143/PTPS.160.1/5161961/160-1.pdf}
  {http://oup.prod.sis.lan/ptps/article-pdf/doi/10.1143/PTPS.160.1/5161961/160-1.pdf}
  \BibitemShut {NoStop}%
\bibitem [{\citenamefont {Yanase}\ and\ \citenamefont
  {Ogata}(2003)}]{YanJPSJ03}%
  \BibitemOpen
  \bibfield  {author} {\bibinfo {author} {\bibfnamefont {Y.}~\bibnamefont
  {Yanase}}\ and\ \bibinfo {author} {\bibfnamefont {M.}~\bibnamefont {Ogata}},\
  }\href {\doibase 10.1143/JPSJ.72.673} {\bibfield  {journal} {\bibinfo
  {journal} {Journal of the Physical Society of Japan}\ }\textbf {\bibinfo
  {volume} {72}},\ \bibinfo {pages} {673} (\bibinfo {year} {2003})},\ \Eprint
  {http://arxiv.org/abs/https://doi.org/10.1143/JPSJ.72.673}
  {https://doi.org/10.1143/JPSJ.72.673} \BibitemShut {NoStop}%
\bibitem [{\citenamefont {Annett}\ \emph {et~al.}(2006)\citenamefont {Annett},
  \citenamefont {Litak}, \citenamefont {Gy\"orffy},\ and\ \citenamefont
  {Wysoki\ifmmode~\acute{n}\else \'{n}\fi{}ski}}]{AnnPRB06}%
  \BibitemOpen
  \bibfield  {author} {\bibinfo {author} {\bibfnamefont {J.~F.}\ \bibnamefont
  {Annett}}, \bibinfo {author} {\bibfnamefont {G.}~\bibnamefont {Litak}},
  \bibinfo {author} {\bibfnamefont {B.~L.}\ \bibnamefont {Gy\"orffy}}, \ and\
  \bibinfo {author} {\bibfnamefont {K.~I.}\ \bibnamefont
  {Wysoki\ifmmode~\acute{n}\else \'{n}\fi{}ski}},\ }\href {\doibase
  10.1103/PhysRevB.73.134501} {\bibfield  {journal} {\bibinfo  {journal} {Phys.
  Rev. B}\ }\textbf {\bibinfo {volume} {73}},\ \bibinfo {pages} {134501}
  (\bibinfo {year} {2006})}\BibitemShut {NoStop}%
\bibitem [{\citenamefont {Ng}\ and\ \citenamefont {Sigrist}(2000)}]{NgEPL00}%
  \BibitemOpen
  \bibfield  {author} {\bibinfo {author} {\bibfnamefont {K.~K.}\ \bibnamefont
  {Ng}}\ and\ \bibinfo {author} {\bibfnamefont {M.}~\bibnamefont {Sigrist}},\
  }\href {\doibase 10.1209/epl/i2000-00173-x} {\bibfield  {journal} {\bibinfo
  {journal} {Europhysics Letters ({EPL})}\ }\textbf {\bibinfo {volume} {49}},\
  \bibinfo {pages} {473} (\bibinfo {year} {2000})}\BibitemShut {NoStop}%
\bibitem [{\citenamefont {{R{\o}mer}}\ \emph {et~al.}(2019)\citenamefont
  {{R{\o}mer}}, \citenamefont {{Scherer}}, \citenamefont {{Eremin}},
  \citenamefont {{Hirschfeld}},\ and\ \citenamefont {{Andersen}}}]{Romer2019}%
  \BibitemOpen
  \bibfield  {author} {\bibinfo {author} {\bibfnamefont {A.~T.}\ \bibnamefont
  {{R{\o}mer}}}, \bibinfo {author} {\bibfnamefont {D.~D.}\ \bibnamefont
  {{Scherer}}}, \bibinfo {author} {\bibfnamefont {I.~M.}\ \bibnamefont
  {{Eremin}}}, \bibinfo {author} {\bibfnamefont {P.~J.}\ \bibnamefont
  {{Hirschfeld}}}, \ and\ \bibinfo {author} {\bibfnamefont {B.~M.}\
  \bibnamefont {{Andersen}}},\ }\href@noop {} {\bibfield  {journal} {\bibinfo
  {journal} {arXiv e-prints}\ ,\ \bibinfo {eid} {arXiv:1905.04782}} (\bibinfo
  {year} {2019})},\ \Eprint {http://arxiv.org/abs/1905.04782} {arXiv:1905.04782
  [cond-mat.supr-con]} \BibitemShut {NoStop}%
\bibitem [{\citenamefont {Haverkort}\ \emph {et~al.}(2008)\citenamefont
  {Haverkort}, \citenamefont {Elfimov}, \citenamefont {Tjeng}, \citenamefont
  {Sawatzky},\ and\ \citenamefont {Damascelli}}]{HavPRL08}%
  \BibitemOpen
  \bibfield  {author} {\bibinfo {author} {\bibfnamefont {M.~W.}\ \bibnamefont
  {Haverkort}}, \bibinfo {author} {\bibfnamefont {I.~S.}\ \bibnamefont
  {Elfimov}}, \bibinfo {author} {\bibfnamefont {L.~H.}\ \bibnamefont {Tjeng}},
  \bibinfo {author} {\bibfnamefont {G.~A.}\ \bibnamefont {Sawatzky}}, \ and\
  \bibinfo {author} {\bibfnamefont {A.}~\bibnamefont {Damascelli}},\ }\href
  {\doibase 10.1103/PhysRevLett.101.026406} {\bibfield  {journal} {\bibinfo
  {journal} {Phys. Rev. Lett.}\ }\textbf {\bibinfo {volume} {101}},\ \bibinfo
  {pages} {026406} (\bibinfo {year} {2008})}\BibitemShut {NoStop}%
\bibitem [{\citenamefont {Veenstra}\ \emph {et~al.}(2014)\citenamefont
  {Veenstra}, \citenamefont {Zhu}, \citenamefont {Raichle}, \citenamefont
  {Ludbrook}, \citenamefont {Nicolaou}, \citenamefont {Slomski}, \citenamefont
  {Landolt}, \citenamefont {Kittaka}, \citenamefont {Maeno}, \citenamefont
  {Dil}, \citenamefont {Elfimov}, \citenamefont {Haverkort},\ and\
  \citenamefont {Damascelli}}]{VeePRL14}%
  \BibitemOpen
  \bibfield  {author} {\bibinfo {author} {\bibfnamefont {C.~N.}\ \bibnamefont
  {Veenstra}}, \bibinfo {author} {\bibfnamefont {Z.-H.}\ \bibnamefont {Zhu}},
  \bibinfo {author} {\bibfnamefont {M.}~\bibnamefont {Raichle}}, \bibinfo
  {author} {\bibfnamefont {B.~M.}\ \bibnamefont {Ludbrook}}, \bibinfo {author}
  {\bibfnamefont {A.}~\bibnamefont {Nicolaou}}, \bibinfo {author}
  {\bibfnamefont {B.}~\bibnamefont {Slomski}}, \bibinfo {author} {\bibfnamefont
  {G.}~\bibnamefont {Landolt}}, \bibinfo {author} {\bibfnamefont
  {S.}~\bibnamefont {Kittaka}}, \bibinfo {author} {\bibfnamefont
  {Y.}~\bibnamefont {Maeno}}, \bibinfo {author} {\bibfnamefont {J.~H.}\
  \bibnamefont {Dil}}, \bibinfo {author} {\bibfnamefont {I.~S.}\ \bibnamefont
  {Elfimov}}, \bibinfo {author} {\bibfnamefont {M.~W.}\ \bibnamefont
  {Haverkort}}, \ and\ \bibinfo {author} {\bibfnamefont {A.}~\bibnamefont
  {Damascelli}},\ }\href {\doibase 10.1103/PhysRevLett.112.127002} {\bibfield
  {journal} {\bibinfo  {journal} {Phys. Rev. Lett.}\ }\textbf {\bibinfo
  {volume} {112}},\ \bibinfo {pages} {127002} (\bibinfo {year}
  {2014})}\BibitemShut {NoStop}%
\bibitem [{\citenamefont {Fatuzzo}\ \emph {et~al.}(2015)\citenamefont
  {Fatuzzo}, \citenamefont {Dantz}, \citenamefont {Fatale}, \citenamefont
  {Olalde-Velasco}, \citenamefont {Shaik}, \citenamefont {Dalla~Piazza},
  \citenamefont {Toth}, \citenamefont {Pelliciari}, \citenamefont {Fittipaldi},
  \citenamefont {Vecchione}, \citenamefont {Kikugawa}, \citenamefont {Brooks},
  \citenamefont {R\o{}nnow}, \citenamefont {Grioni}, \citenamefont {R\"uegg},
  \citenamefont {Schmitt},\ and\ \citenamefont {Chang}}]{FatPRB15}%
  \BibitemOpen
  \bibfield  {author} {\bibinfo {author} {\bibfnamefont {C.~G.}\ \bibnamefont
  {Fatuzzo}}, \bibinfo {author} {\bibfnamefont {M.}~\bibnamefont {Dantz}},
  \bibinfo {author} {\bibfnamefont {S.}~\bibnamefont {Fatale}}, \bibinfo
  {author} {\bibfnamefont {P.}~\bibnamefont {Olalde-Velasco}}, \bibinfo
  {author} {\bibfnamefont {N.~E.}\ \bibnamefont {Shaik}}, \bibinfo {author}
  {\bibfnamefont {B.}~\bibnamefont {Dalla~Piazza}}, \bibinfo {author}
  {\bibfnamefont {S.}~\bibnamefont {Toth}}, \bibinfo {author} {\bibfnamefont
  {J.}~\bibnamefont {Pelliciari}}, \bibinfo {author} {\bibfnamefont
  {R.}~\bibnamefont {Fittipaldi}}, \bibinfo {author} {\bibfnamefont
  {A.}~\bibnamefont {Vecchione}}, \bibinfo {author} {\bibfnamefont
  {N.}~\bibnamefont {Kikugawa}}, \bibinfo {author} {\bibfnamefont {J.~S.}\
  \bibnamefont {Brooks}}, \bibinfo {author} {\bibfnamefont {H.~M.}\
  \bibnamefont {R\o{}nnow}}, \bibinfo {author} {\bibfnamefont {M.}~\bibnamefont
  {Grioni}}, \bibinfo {author} {\bibfnamefont {C.}~\bibnamefont {R\"uegg}},
  \bibinfo {author} {\bibfnamefont {T.}~\bibnamefont {Schmitt}}, \ and\
  \bibinfo {author} {\bibfnamefont {J.}~\bibnamefont {Chang}},\ }\href
  {\doibase 10.1103/PhysRevB.91.155104} {\bibfield  {journal} {\bibinfo
  {journal} {Phys. Rev. B}\ }\textbf {\bibinfo {volume} {91}},\ \bibinfo
  {pages} {155104} (\bibinfo {year} {2015})}\BibitemShut {NoStop}%
\bibitem [{\citenamefont {Zhitomirsky}\ and\ \citenamefont
  {Rice}(2001)}]{ZhiPRL01}%
  \BibitemOpen
  \bibfield  {author} {\bibinfo {author} {\bibfnamefont {M.~E.}\ \bibnamefont
  {Zhitomirsky}}\ and\ \bibinfo {author} {\bibfnamefont {T.~M.}\ \bibnamefont
  {Rice}},\ }\href {\doibase 10.1103/PhysRevLett.87.057001} {\bibfield
  {journal} {\bibinfo  {journal} {Phys. Rev. Lett.}\ }\textbf {\bibinfo
  {volume} {87}},\ \bibinfo {pages} {057001} (\bibinfo {year}
  {2001})}\BibitemShut {NoStop}%
\bibitem [{\citenamefont {Annett}\ \emph {et~al.}(2002)\citenamefont {Annett},
  \citenamefont {Litak}, \citenamefont {Gy\"orffy},\ and\ \citenamefont
  {Wysoki\ifmmode~\acute{n}\else \'{n}\fi{}ski}}]{AnnPRB02}%
  \BibitemOpen
  \bibfield  {author} {\bibinfo {author} {\bibfnamefont {J.~F.}\ \bibnamefont
  {Annett}}, \bibinfo {author} {\bibfnamefont {G.}~\bibnamefont {Litak}},
  \bibinfo {author} {\bibfnamefont {B.~L.}\ \bibnamefont {Gy\"orffy}}, \ and\
  \bibinfo {author} {\bibfnamefont {K.~I.}\ \bibnamefont
  {Wysoki\ifmmode~\acute{n}\else \'{n}\fi{}ski}},\ }\href {\doibase
  10.1103/PhysRevB.66.134514} {\bibfield  {journal} {\bibinfo  {journal} {Phys.
  Rev. B}\ }\textbf {\bibinfo {volume} {66}},\ \bibinfo {pages} {134514}
  (\bibinfo {year} {2002})}\BibitemShut {NoStop}%
\bibitem [{\citenamefont {Huang}\ and\ \citenamefont {Yao}(2018)}]{WenPRL18}%
  \BibitemOpen
  \bibfield  {author} {\bibinfo {author} {\bibfnamefont {W.}~\bibnamefont
  {Huang}}\ and\ \bibinfo {author} {\bibfnamefont {H.}~\bibnamefont {Yao}},\
  }\href {\doibase 10.1103/PhysRevLett.121.157002} {\bibfield  {journal}
  {\bibinfo  {journal} {Phys. Rev. Lett.}\ }\textbf {\bibinfo {volume} {121}},\
  \bibinfo {pages} {157002} (\bibinfo {year} {2018})}\BibitemShut {NoStop}%
\bibitem [{\citenamefont {R\o{}ising}\ \emph {et~al.}(2018)\citenamefont
  {R\o{}ising}, \citenamefont {Flicker}, \citenamefont {Scaffidi},\ and\
  \citenamefont {Simon}}]{RoisingPRB2018}%
  \BibitemOpen
  \bibfield  {author} {\bibinfo {author} {\bibfnamefont {H.~S.}\ \bibnamefont
  {R\o{}ising}}, \bibinfo {author} {\bibfnamefont {F.}~\bibnamefont {Flicker}},
  \bibinfo {author} {\bibfnamefont {T.}~\bibnamefont {Scaffidi}}, \ and\
  \bibinfo {author} {\bibfnamefont {S.~H.}\ \bibnamefont {Simon}},\ }\href
  {\doibase 10.1103/PhysRevB.98.224515} {\bibfield  {journal} {\bibinfo
  {journal} {Phys. Rev. B}\ }\textbf {\bibinfo {volume} {98}},\ \bibinfo
  {pages} {224515} (\bibinfo {year} {2018})}\BibitemShut {NoStop}%
\bibitem [{\citenamefont {{Huang}}\ \emph {et~al.}(2019)\citenamefont
  {{Huang}}, \citenamefont {{Zhou}},\ and\ \citenamefont
  {{Yao}}}]{HuangArxiv19}%
  \BibitemOpen
  \bibfield  {author} {\bibinfo {author} {\bibfnamefont {W.}~\bibnamefont
  {{Huang}}}, \bibinfo {author} {\bibfnamefont {Y.}~\bibnamefont {{Zhou}}}, \
  and\ \bibinfo {author} {\bibfnamefont {H.}~\bibnamefont {{Yao}}},\
  }\href@noop {} {\bibfield  {journal} {\bibinfo  {journal} {arXiv e-prints}\
  ,\ \bibinfo {eid} {arXiv:1905.03523}} (\bibinfo {year} {2019})},\ \Eprint
  {http://arxiv.org/abs/1905.03523} {arXiv:1905.03523 [cond-mat.supr-con]}
  \BibitemShut {NoStop}%
\bibitem [{\citenamefont {{Gingras}}\ \emph {et~al.}(2018)\citenamefont
  {{Gingras}}, \citenamefont {{Nourafkan}}, \citenamefont {{Tremblay}},\ and\
  \citenamefont {{C{\^o}t{\'e}}}}]{GingrasArxiv18}%
  \BibitemOpen
  \bibfield  {author} {\bibinfo {author} {\bibfnamefont {O.}~\bibnamefont
  {{Gingras}}}, \bibinfo {author} {\bibfnamefont {R.}~\bibnamefont
  {{Nourafkan}}}, \bibinfo {author} {\bibfnamefont {A.-M.~S.}\ \bibnamefont
  {{Tremblay}}}, \ and\ \bibinfo {author} {\bibfnamefont {M.}~\bibnamefont
  {{C{\^o}t{\'e}}}},\ }\href@noop {} {\bibfield  {journal} {\bibinfo  {journal}
  {arXiv e-prints}\ ,\ \bibinfo {eid} {arXiv:1808.02527}} (\bibinfo {year}
  {2018})},\ \Eprint {http://arxiv.org/abs/1808.02527} {arXiv:1808.02527
  [cond-mat.supr-con]} \BibitemShut {NoStop}%
\bibitem [{\citenamefont {Cuoco}\ \emph {et~al.}(1998)\citenamefont {Cuoco},
  \citenamefont {Noce},\ and\ \citenamefont {Romano}}]{CuocoPRB98}%
  \BibitemOpen
  \bibfield  {author} {\bibinfo {author} {\bibfnamefont {M.}~\bibnamefont
  {Cuoco}}, \bibinfo {author} {\bibfnamefont {C.}~\bibnamefont {Noce}}, \ and\
  \bibinfo {author} {\bibfnamefont {A.}~\bibnamefont {Romano}},\ }\href
  {\doibase 10.1103/PhysRevB.57.11989} {\bibfield  {journal} {\bibinfo
  {journal} {Phys. Rev. B}\ }\textbf {\bibinfo {volume} {57}},\ \bibinfo
  {pages} {11989} (\bibinfo {year} {1998})}\BibitemShut {NoStop}%
\bibitem [{\citenamefont {{Strand}}\ \emph {et~al.}(2019)\citenamefont
  {{Strand}}, \citenamefont {{Zingl}}, \citenamefont {{Wentzell}},
  \citenamefont {{Parcollet}},\ and\ \citenamefont {{Georges}}}]{StraArxiv19}%
  \BibitemOpen
  \bibfield  {author} {\bibinfo {author} {\bibfnamefont {H.~U.~R.}\
  \bibnamefont {{Strand}}}, \bibinfo {author} {\bibfnamefont {M.}~\bibnamefont
  {{Zingl}}}, \bibinfo {author} {\bibfnamefont {N.}~\bibnamefont {{Wentzell}}},
  \bibinfo {author} {\bibfnamefont {O.}~\bibnamefont {{Parcollet}}}, \ and\
  \bibinfo {author} {\bibfnamefont {A.}~\bibnamefont {{Georges}}},\ }\href@noop
  {} {\bibfield  {journal} {\bibinfo  {journal} {arXiv e-prints}\ ,\ \bibinfo
  {eid} {arXiv:1904.07324}} (\bibinfo {year} {2019})},\ \Eprint
  {http://arxiv.org/abs/1904.07324} {arXiv:1904.07324 [cond-mat.str-el]}
  \BibitemShut {NoStop}%
\end{thebibliography}%


\end{document}